\documentclass[useAMS,usenatbib]{mn2e}
\pdfoutput=1
\usepackage{aas_macros}
\usepackage{epsfig}
\usepackage{amsmath}
\usepackage{amssymb}
\usepackage{caption}
\usepackage{leftidx}
\usepackage{natbib}
\usepackage{float}

\usepackage[T1]{fontenc}
\usepackage{lmodern}
\usepackage[rgb]{xcolor}
\usepackage{listings}
\definecolor{MyGreen}{rgb}{0.0,0.6,0.3}
\definecolor{MyPurple}{rgb}{0.6,0,0.3}
\usepackage{fancyref}
\usepackage{hyperref}
\hypersetup{colorlinks=true,citecolor=MyGreen,linkcolor=MyPurple,urlcolor=blue}

\def\beq{\begin{equation}}
\def\eeq{\end{equation}}
\def\ba{\begin{eqnarray}}
\def\ea{\end{eqnarray}}
\def\bal{\begin{align}}
\def\eal{\end{align}}
\def\bxi{{\mbox{\boldmath $\xi$}}}

\begin{document}

\title[Pulsar Spins] {Angular momentum transport in massive stars and natal neutron star rotation rates}

\author[Ma \& Fuller]{
Linhao Ma$^{1, 2, }$\thanks{Email: fluorine@mail.ustc.edu.cn} and 
Jim Fuller$^{2}$
\\$^1$Department of Modern Physics, University of Science and Technology of China, Hefei, Anhui 230026, China
\\$^2$TAPIR, Mailcode 350-17, California Institute of Technology, Pasadena, CA 91125, USA
}

\label{firstpage}
\maketitle

\begin{abstract}

The internal rotational dynamics of massive stars are poorly understood. If angular momentum (AM) transport between the core and the envelope is inefficient, the large core AM upon core-collapse will produce rapidly rotating neutron stars (NSs). However, observations of low-mass stars suggest an efficient AM transport mechanism is at work, which could drastically reduce NS spin rates. Here we study the effects of the baroclinic instability and the magnetic Tayler instability in differentially rotating radiative zones. Although the baroclinic instability may occur, the Tayler instability is likely to be more effective for AM transport. We implement Tayler torques as prescribed by \citealt{fuller:19} into models of massive stars, finding they remove the vast majority of the core's AM as it contracts between the main sequence and helium-burning phases of evolution. If core AM is conserved during core-collapse, we predict natal NS rotation periods of $P_{\rm NS} \approx 50-200 \, {\rm ms}$, suggesting these torques help explain the relatively slow rotation rates of most young NSs, and the rarity of rapidly rotating engine-driven supernovae. Stochastic spin-up via waves just before core-collapse, asymmetric explosions, and various binary evolution scenarios may increase the initial rotation rates of many NSs.

\end{abstract}


\section{Introduction}

Internal rotation rates of massive stars are weakly constrained and poorly understood. While we have a basic understanding of massive stellar evolution and the compact objects that are produced upon core-collapse, we do not even have a zeroth order understanding of their internal rotation rates. This ignorance prevents the development of a detailed understanding of massive star evolution, as rotational effects (e.g., rotational mixing, \citealt{Maeder_2000}) can have a substantial impact on a star's evolution. More importantly, the core rotation rate of a massive star is crucial for determining the outcome of core-collapse. A variety of transients such as gamma-ray bursts (GRBS, \citealt{woosley:93,macfadyen:99,metzger:11}) and superluminous supernovae \citep{kasen:09,nicholl:17} are thought to be engine-powered events that tap into the huge rotational energy reservoir ($E_{\rm rot} \gtrsim 10^{52}$ erg) that can be supplied by a rapidly rotating neutron star (NS) or black hole (BH). Theoretical predictions for core rotation rates depend on the unknown efficiency of AM transport. Zero AM transport implies that nearly all compact objects will be born maximally rotating \citep{heger:00}, while instantaneous AM tranpsort (i.e., rigid stellar rotation) implies compact objects will be essentially non-rotating. Both models conflict with observations, indicating that strong but imperfect AM coupling mechanisms are at work.

Observational constraints on core rotation arise primarily from compact object rotation rates. While some young neutron stars (e.g., the Crab pulsar) have natal spin periods of $P_0 \sim 20 \, {\rm ms}$ \citep{kaspi:02}, the majority of NSs have longer natal spin periods of $P_0 \sim 50-100 \, {\rm ms}$ \citep{faucher:06,popov:10,popov:12,gullon:14}, and some young NSs have natal spin periods as long as $P_0 \sim 400 {\rm ms}$ \citep{gotthelf:13}. Spin rates of accreting BHs in X-ray binaries (XRBs) can be estimated from accretion disk modeling, and current estimates suggest a broad range of spin-rates ($0.1 \lesssim a \lesssim 1$) \citep{miller:15}. Many of these spin rates may have been increased by post-BH formation mass transfer, while some may reflect natal spin rates, but these can be affected by binary interactions \citep{valsecchi:10,qin:18,qin:18b}. BH mergers detected by LIGO \citep{ligorun1,ligo170104,ligo170608,ligo170814} are now providing the first spin rates of non-accreting BHs, and the low values of $\chi_{\rm eff}$ measured so far may indicate low BH natal spins \citep{ligoo2b:18,zaldarriaga:18}.

A crucial piece of observational evidence arises from asteroseismic rotation rates measured in {\it low-mass} ($M \lesssim 3 \, M_\odot$) stars. These measurements have been performed for main-sequence, red-giant branch, horizontal branch, and white dwarf stars \citep{beck:12,mosser:12,deheuvels:14,deheuvels:15,benomar:15,hermes:17,gehan:18}. Their message is clear: stellar cores and compact objects rotate orders of magnitude slower than they would in the absence of AM transport, and the rotation rates are slower than predicted by most AM transport models \citep{cantiello:14}. Hence, massive stellar cores may also rotate slower than prior predictions. Perhaps the most realistic predictions come from \cite{heger:05} and \cite{wheeler:15} upon inclusion of the Tayler-Spruit dynamo \citep{spruit:02} and/or magneto-rotational instability (MRI) into stellar models, and they predict typical NS rotation rates $P_{\rm NS} \sim 10 \, {\rm ms}$, at the fast end of the distribution inferred from radio pulsars. However, the same physics implemented in low-mass stars predict rotation rates roughly an order of magnitude faster than observed in red giant cores and white dwarfs \citep{cantiello:14}.

Recently, \cite{fuller:19} argued that the Tayler instability (e.g., \citealt{spruit:99}) can grow to larger saturated amplitudes and provide more efficient AM than predicted by \cite{spruit:02}, and they provided an updated prescription for the ``Tayler torques" arising from the instability. Models incorporating this prescription produced a good match with asteroseismic measurements of stars on the main sequence, red giant branch, red clump, and white dwarf cooling track. In this work, we extend the models of \cite{fuller:19} to the massive star regime ($M \gtrsim 10 \, M_\odot)$, and we predict the AM contained in the core as it approaches core-collapse, and the corresponding rotation rate of a NS or BH if AM is conserved during collapse. Incorporating Tayler torques into massive stellar models, we find significantly slower core rotation rates (by roughly one order of magnitude) than prior predictions. We also examine the previously neglected baroclinic instability, and we provide a more rigorous derivation of the dispersion relation for Tayler instabilities.

In Sections \ref{sec2} and \ref{sec3}, we investigate the baroclinic and Tayler instabilities. Section \ref{sec4} describes our stellar models and AM transport results. In Section \ref{sec5}, we discuss the implications of our results for massive stellar evolution and massive stellar death, and we conclude in Section \ref{sec6}. Most of the detailed calculations regarding baroclinic and Tayler instabilities are presented in the appendices.

In what follows we refer to many variables whose definitions can be found in Table \ref{tab:my_label}.

\section{Baroclinic Instability}
\label{sec2}

A well-known hydrodynamic instability that might be important for AM transport is the baroclinic instability, which has only been examined in a few previous astrophysical works \citep[e.g.,][]{tassoul:82,fujimoto:87,fujimoto:88,kitchatinov:14}, but has been extensively studied in Earth atmospheric and oceanic contexts \citep[see review in][]{pedlosky:92}. Except in the case of a cylindrical rotation profile, differential rotation generally displaces the surfaces of constant density from isobars, creating a baroclinic stellar structure with a density/entropy gradient along isobars. The baroclinic instability is sourced by the potential energy released by latitudinal exchange between high and low density fluid elements, and is analogous to convection in the horizontal direction driven by the latitudinal entropy gradient. Despite its ubiquity, the baroclinic instability has frequently been neglected in astrophysics due to a series of papers claiming that it does not usually occur in stars \citep{,knobloch:82,knobloch:83,spruit:83,spruit:84,zahn:93}. Below we show why the conclusion of these works are mostly erroneous, but why baroclinic instabilities often remain unimportant.

Under the assumption of an axisymmetric background, baroclinicity is related to differential rotation by \citep{kitchatinov:14}
\beq
\label{baro1}
r\sin\theta\bigg(\cos\theta\frac{\partial}{\partial r}-\frac{\sin\theta}{r}\frac{\partial}{\partial \theta}\bigg)\Omega^2=-\frac{1}{\rho^2}(\nabla\rho\times\nabla P)_\phi
\eeq
where $(r,\theta,\phi)$ are spherical coordinates, $\Omega$ is the angular frequency, $\rho$ is the density and $P$ is the pressure. In our analysis we ignore the effect of centrifugal distortion, such that $\nabla P$ is in the radial direction (accounting for centrifugal distortion, one can simply redefine the radial coordinate to be perpendicular to isobars). Baroclinic instability occurs when $\partial \rho/\partial \theta\neq0$ which can occur when there is differential rotation. We define
\beq
N_\theta^2\equiv-\frac{g}{\rho r}\frac{\partial \rho}{\partial \theta}
\eeq
to be the characteristic buoyancy frequency for baroclinic instability, where $g$ is the gravity. In our analysis we shall assume shellular rotation, ($\partial \Omega /\partial \theta=0$), such that equation \eqref{baro1} reduces to
\beq
N_\theta^2=2q\Omega^2\sin\theta\cos\theta \, ,
\eeq
where 
\beq
q=\frac{d \ln \Omega}{d \ln r}
\eeq
is the dimensionless shear. 

We start by analyzing the stability of oscillation modes, assuming the time dependence of each perturbation variable $\delta Q$ is
\beq
\label{osc}
\delta Q\propto e^{-i\omega t} \, .
\eeq
where $\omega$ is the complex oscillation frequency. In the rotating frame, we have the perturbed momentum equation
\beq
\label{eqM}
-\omega^2 \vec{\xi}=-\frac{1}{\rho}\nabla\delta P-\frac{g}{\rho}\delta \rho \, \hat{r} + 2i\omega\vec{\Omega}\times\vec{\xi}
\eeq
where $\vec{\xi}$ is the perturbed displacement and $\hat{r}$ is the unit vector in the radius direction, and $\delta P$ and $\delta \rho$ are the Eulerian pressure and density perturbations. We have made the Cowling approximation by neglecting the perturbation to the gravitational acceleration. We also have the perturbed continuity equation
\beq
\label{eqC}
\delta \rho+\nabla\cdot(\rho \vec{\xi})=0 \, .
\eeq
We show in appendix \ref{appendixa} that the energy equation with the baroclinic term and thermal diffusion is
\beq
\label{eqE}
\frac{\delta \rho}{\rho}=\frac{1}{\tilde{c_\mathrm{s}}^2}\frac{\delta P}{\rho}+\xi_r\frac{\tilde{N}^2}{g}+\xi_\theta\frac{\tilde{N_\theta}^2}{g}
\eeq
with all the notations made clear in appendix \ref{appendixa}.

The set of equations \eqref{eqM}, \eqref{eqC} and \eqref{eqE} are difficult to solve, even for linear theory, because the Coriolis and baroclinic terms break the spherical symmetry of the problem. The growing modes of interest are often low frequency and nearly incompressible gravity modes and Rossby modes, so here we consider two approximations that are often used for such modes, namely, the traditional approximation and the geostrophic approximation.

\subsection{Traditional Approximation}

Here we perform a linear stability analysis for modes computed using the traditional approximation (e.g., \citealt{bildsten:96,lee:97}), decomposing the spatial dependence of the mode as 
\beq
\delta Q\propto \delta Q(\theta) \exp \bigg[ i\bigg(\int k_r dr+m\phi-\omega t\bigg) \bigg] \, .
\eeq
We perform a local analysis and make a WKB approximation $k_r \gg 1/H, 1/r$, since the radial wavenumbers of low frequency modes are large due to the strong buoyancy force.

We show in appendix \ref{appendixb1} that when thermal diffusion is small (such that $k^2 \kappa \ll \omega$), this approximation leads to the growth rate
\beq
\label{tradgr}
\gamma=\mathrm{Im}(\omega)=-\frac{1}{2}\frac{N_T^2}{N^2}\kappa k_r^2\pm \frac{1}{2}\mathrm{sign}(k_r)\frac{N_\theta^2 F}{(N^2\lambda)^\frac{1}{2}}
\eeq
where $N$ is the Brunt-V\"ais\"al\"a frequency, $N_T$ is  its thermal component, $\kappa$ is the thermal diffusivity, $\lambda$ is the horizontal eigenvalue of a Hough function, and $F$ is an angular overlap integral (see appendix \ref{appendixb1} for details). The first term in equation \ref{tradgr} is due to thermal diffusion and always causes damping in radiative regions with $N_T^2$ and $N^2$ positive. The second term is due to baroclinicity and causes damping or driving depending on the sign of $k_r$. Hence, only modes propagating in one direction will be driven. In the absence of thermal diffusion, typical growth rates are thus $\gamma \sim q \Omega^2/N$, remarkably independent of radial wavenumber $k_r$ (though with some dependence on horizontal wavenumber through the $F$ and $\lambda$ terms). This explains the numerical results of \cite{kitchatinov:14}, who found a nearly constant growth rate as a function of $k_r$, except at large $k_r$ where thermal diffusion is important. The transition to damped modes occurs when  $N N_\theta^2 \lesssim N_T^2\kappa k_r^2$.

We pause to reconcile the result of equation \ref{tradgr} with previous works such as \cite{spruit:83} and \cite{zahn:93}, which claim that baroclinic instability requires very large shear and is unlikely to occur in stars, even in the absence of diffusive effects. Prior claims relied on a theorem \citep{charney:62} from geophysics based on potential vorticity. This theorem involves global analysis, whereas equation \ref{tradgr} is derived based on a local analysis. If one integrates equation \ref{tradgr} for a standing mode to determine its global growth rate, the baroclinic growth term is equal to zero, because the inwardly and outwardly propagating component of the standing mode (i.e., the positive and negative signs of $k_r$) cancel each other out. Only non-WKB terms contribute to the global growth rate, and these terms require large shear for growth (they typically require $q \Omega^2 \gtrsim N^2$, a criterion similar to the Richardson criterion for instability of shear flows). 

Nonetheless, the theorem from \cite{charney:62} is not generally applicable, as is well-known in geophysics (see review in \citealt{pedlosky:92}), and it should not be considered a necessary condition for baroclinic instability. The subtle reason is that the potential vorticity argument is based on a global calculation in which boundary terms can be substituted after assuming no divergences within an integration domain. However, such divergences are ubiquitous at critical layers where the local frequency of a mode goes to zero, and its wavenumber diverges. The critical layers allow a growing mode propagating in one direction to be absorbed without propagating back to become damped.  For this reason, growing baroclinic modes are ubiquitous in atmoshperic contexts and may generally exist in stars as well.

However, because critical layers are required for the excitation of baroclinic modes, only modes of low enough frequency can be excited. Consider a mode whose local frequency (measured in a corotating frame) at a critical layer is $\omega(r_c) = 0$. Then the mode's local frequency at a different layer in the star, whose angular rotation rate is different by $\Delta \Omega$, is $\omega(r) = m \Delta \Omega$. Using the g-mode dispersion relation
\beq
k_r^2 \simeq \frac{\lambda N^2}{r^2 \omega^2} \, 
\eeq
we find that mode growth requires
$N_\theta^2 (m \Delta \Omega)^2 r^2 \gtrsim \lambda^{3/2} N_T^2 N \kappa \,$. The maximum possible value of $\Delta \Omega$ is simply the maximal rotation rate at any point in the stars, $\Omega_{\rm max}$. Since $m \leq \ell \lesssim \lambda^{1/2}$, and $N_\theta^2 \sim q \Omega^2$ baroclinic instability requires
\beq
\label{barinst}
q \Omega^2 \Omega_{\rm max}^2 \gtrsim N_T^2 N \frac{\kappa}{r^2} \, .
\eeq
In Sun-like stars and red giants, $\Omega/N$ is typically of order $10^{-3}$ or smaller. In many situations, we find that equation \ref{barinst} is not satisfied anywhere in the radiative regions of stars, so critical layers will not exist for modes with frequencies high enough to avoid radiative damping. For such standing modes, baroclinic growth will typically not occur, so the baroclinic instability will not operate. An important exception is rapidly rotating stars, and stars with strong shear layers near the surface due to ongoing accretion (e.g., \citealt{piro:07}), where the baroclinic instability is more likely to occur. 

When the thermal diffusion is large, we take the limit $\kappa\rightarrow\infty$, and show in Appendix \ref{appendixb1} that the growth rate reduces to
\beq
\gamma=\mathrm{Im}(\omega)=-\frac{N_T^2\lambda}{2\kappa k_r^4 r^2} \, .
\eeq
This is always a damping mode so the instability does not exist at large thermal diffusivity. However, our analysis does not include compositional diffusivity or viscosity. Interestingly, when included, they give rise to new classes of instabilities such as the ABCD instability \citep{spruit:83}. We find it unlikely that such instabilities contribute significantly to AM transport because they vastly over predict the differential rotation in the radiative core of the Sun.

\subsection{Geostrophic Approximation}

\begin{figure}
\begin{center}
\includegraphics[scale=0.5]{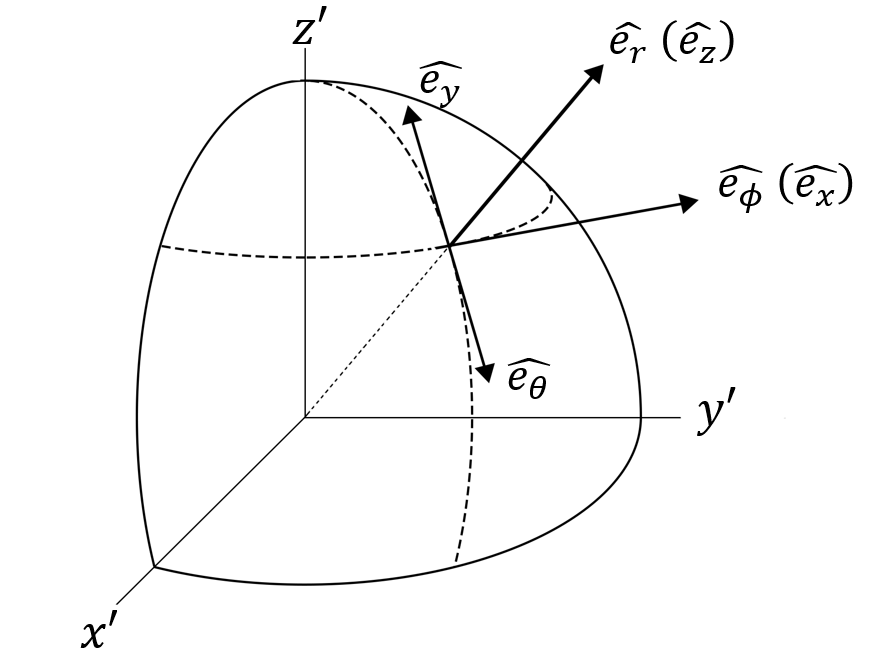}
\end{center} 
\caption{\label{beta-plane} The relation between local Cartesian coordinates and spherical coordinates used for the geostrophic approximation. The original Cartesian coordinates are labeled with $(x',y',z')$.}
\end{figure}

An alternative method to study the problem locally is to apply the geostrophic approximation (e.g., \citealt{spruit:84}). To do this, we restrict ourselves to low-frequency modes and neglect higher order terms of $\omega$. We perform the analysis in a thin box near a given point $(r_0,\theta_0,\phi_0)$, establishing a local coordinate frame $(x,y,z)$ with the unit vectors in their directions $(\hat{e}_x,\hat{e}_y,\hat{e}_z)$ corresponding to the original unit vectors $(\hat{e}_\phi,-\hat{e}_\theta,\hat{e}_r)$ respectively (Figure \ref{beta-plane}). Then within the box we have
\begin{align}
x&=r_0\sin\theta_0(\phi-\phi_0) \, ,\\
y&=-r_0(\theta-\theta_0) \, , \\
z&=r-r_0 \, .
\end{align}

For local modes, we can express the perturbation variables as
\beq
\delta Q\propto\exp \big[ i(k_x x+k_y y+k_zz-\omega t) \big] \, .
\eeq
Since we neglect the high order terms of $\omega$, the most significant restoring term in the horizontal direction is the Coriolis force, so we define a Coriolis frequency $f$ as
\beq
f\equiv 2\Omega\cos\theta \, .
\eeq
This is known as the beta-plane approximation\footnote
{
The name is from the definition of the Rossby parameter $\beta$
\begin{equation*}
\beta\equiv\frac{\partial f}{\partial y}=\frac{2\Omega\sin\theta}{r_0} \, .
\end{equation*}
In our `thin box' $f$ varies linearly with the polar angle, i.e. $f=f_0+\beta y$.
}
in geophysical fluid dynamics. (Note that in geophysics the latitude $\varphi$ is often used instead of the polar angle $\theta$ we use here, so the Coriolis parameter can be equivalently defined as $f=2\Omega\sin\varphi$.) Due to the strong stratification in the radial direction, the radial displacements are expected to be much smaller than the horizontal ones, and the radial component of the Coriolis force can be neglected. Then the equation of motion \eqref{eqM} reduces to
\begin{align}
fv_y=\frac{1}{\rho}\frac{\partial}{\partial x}\delta P \, ,\\
fv_x=-\frac{1}{\rho}\frac{\partial}{\partial y}\delta P \, ,\\
g\delta \rho=-\frac{\partial}{\partial z}\delta P \, ,
\end{align}
We also use the incompressible approximation
\beq
\frac{\partial v_x}{\partial x}+\frac{\partial v_y}{\partial y}+\frac{\partial v_z}{\partial z} = 0 \, .
\eeq
When the thermal diffusion is small but not negligible, we show in Appendix \ref{appendixb2} that, with the energy equation, the growth rate is 
\beq
\label{gammageo}
\gamma=\mathrm{Im}(\omega)=-\frac{N_T^2}{N^2}\kappa k^2+\frac{N_\theta^2}{2\Omega\cos\theta}\frac{k_x}{k_z} \, .
\eeq
where $k^2=k_x^2+k_y^2+k_z^2$ is the total wave number. This is similar to the results we found for the traditional approximation, but notably different because the growth rate now depends on both the sign and the value of $k_z$. The reason is the different dispersion relation between the gravito-inertial modes studied in the traditional approximation, and the Rossby modes studied in the geostrophic approximation. The growth rate of equation \ref{gammageo} is generally smaller than that of equation \ref{tradgr} by a factor $\sim k_x/k_z \ll 1$. Hence, thermal diffusion will also typically suppress the growth of Rossby modes growing via the baroclinic instability.

\section{Tayler Instability}
\label{sec3}

Due to their complexity, magnetohydrodynamic instabilities in stellar interiors have received less attention than hydrodynamic instabilities, but they can be more important for AM transport for two reasons. First, they are less quenched by diffusion because the magnetic diffusivity $\eta$ is much smaller than the thermal diffusivity $\kappa$ in stellar interiors. Second, hydrodynamic instabilities can only transport AM via a Reynolds stress $T \sim 4 \pi \rho \langle v_r v_\phi \rangle$ that requires radial motion, which is suppressed by large buoyancy forces. In contrast, magnetic instabilities can produce a Maxwell stress $T \sim B_r B_\phi$ that does not necessarily require radial motion. As shown by \cite{spruit:02} , the magnetic Tayler instability \citep{tayler:73,spruit:99} could play an especially important role in AM transport process.

Prior studies of the Tayler instability have been restricted to a polar region of the star where the geometry becomes Cartesian (e.g., \citealt{acheson:78,spruit:99,zahn:07}). \cite{denissenkov:07} argued that the effective horizontal length scale of the instability must be much smaller than $r$ when one considers the spherical geometry of a star.
\footnote{The argument by \citealt{denissenkov:07} confuses the perturbation wavelength $\ell_r \sim k_r^{-1}$ with the Lagrangian displacement $\xi_r$, and it is invalid for constraining the vertical wavelength $\ell_r \sim \omega_{\rm A} \ell_h/N$ or horizontal wavelength $\ell_h \sim r$. The arguments of \cite{denissenkov:07} are essentially arguments about non-linear geometric and Coriolis terms, which do not affect the linear instability, but may affect its non-linear evolution and saturation.}
Here we present the dispersion relation for Tayler modes at an arbitrary latitude in the star. We will find that the dispersion relation is unaltered apart from some factors of $\cos^2 \theta$, and that order of magnitude estimates based on a polar analysis are valid.

We start from local oscillation analysis, assuming the perturbation variables $\delta Q$ to have dependence 
\beq
\delta Q \propto \exp \big[ i (k_rr+m\phi+l\theta-\omega t) \big] \, .
\eeq
We neglect viscosity, but include thermal and Ohmic diffusion. We apply the same energy equation as above (though we now neglect baroclinicity), and make an incompressible approximation in the sense that sound speed $c_\mathrm{s}\rightarrow\infty$. The eigenvalue of Laplacian can be safely approximated as $k_r^2$ by WKB approximation, hence the energy equation reduces to
\beq
g\frac{\delta\rho}{\rho}=\bigg[N_T^2\bigg(1+i\frac{\kappa k_r^2}{\omega}\bigg)^{-1}+N_\mu^2 \bigg] \xi_r \, ,
\eeq
where $\kappa$ is the thermal diffusivity, and $N_\mu$ is the composition part of Brunt-V\"ais\"al\"a frequency. Note that thermal diffusion suppresses the thermal component of the Brunt-V\"ais\"al\"a frequency, such that the compositional component is often more important in stars. From the linear MHD equations, we have the perturbed induction equation
\beq
-i\omega \vec{b}=\nabla\times(-i\omega \vec{\xi}\times\vec{B})-\eta k_r^2 \vec{b} \, ,
\eeq
where $\vec{B}$ and $\vec{b}$ are the background and perturbed magnetic fields. We also have the perturbed equation of motion
\beq
-\omega^2\vec{\xi} = - \frac{1}{\rho}\nabla\delta P - g\frac{\delta \rho}{\rho} + 2i\omega \Omega\hat{z}\times\vec{\xi} + \vec{L} = 0 \, ,
\eeq
where $\vec{L}$ is the perturbed Lorentz force and $\hat{z}$ is the unit vector in $z$ direction. We show in Appendix \ref{appendixc} that
\begin{align}
L_r&=\frac{m\omega_\mathrm{A}^2}{1+i\eta k_r^2/\omega}(k_rr\sin\theta\xi_\phi-i\sin\theta\xi_\phi-m\xi_r) \, ,\\
L_\theta&=\frac{m\omega_\mathrm{A}^2}{1+i\eta k_r^2/\omega}(l\sin\theta\xi_\phi-m\xi_\theta-2i\cos\theta\xi_\phi) \, ,\\
L_\phi&=\frac{m\omega_\mathrm{A}^2}{1+i\eta k_r^2/\omega}(2i\cos\theta\xi_\theta+i\sin\theta\xi_r) \, ,
\end{align}
where 
\beq
\omega_\mathrm{A} = \frac{B_\phi}{\sqrt{4\pi\rho r^2}}
\eeq
is the Alfv\'{e}n frequency.

We show in appendix \ref{appendixc1} (with all the notations below made clear) that with some assumptions on the background field, these equations lead to the dispersion relation
\beq
\label{taylerdisp}
\begin{split}
&\tilde{\omega}^6-\tilde{\omega}^4(4\tilde{\Omega}^2\cos^2\theta+A_t+A_\mu+2m^2+2hk+h^2)\\
-&\tilde{\omega}^3\cdot 8\tilde{\Omega}m+\tilde{\omega}^2 \big[ m^2A_t+m^2A_\mu+m^2-4m^2\cos^2\theta\\
+&2hk(4\tilde{\Omega}^2\cos^2\theta+A_\mu+m^2)+h^2(4\tilde{\Omega}^2\cos^2\theta+A_t+A_\mu) \big]\\
+&\tilde{\omega}\cdot8mkh\tilde{\Omega}\cos^2\theta-hkm^2A_\mu\\
+&i\tilde{\omega}^5(2h+k)-i\tilde{\omega}^3 \big[ k(4\tilde{\Omega}^2\cos^2\theta+A_\mu+2m^2)\\
+&2h(4\tilde{\Omega}^2\cos^2\theta+A_t+A_\mu+m^2)+h^2k \big] \\
-&i\tilde{\omega}^2\cdot 8(k+h)\tilde{\Omega}\cos^2\theta m\\
+&i\tilde{\omega} \big[ k(m^2A_\mu+m^2-4m^2\cos^2\theta)\\
+&hm^2(A_t+A_\mu)+h^2k(4\tilde{\Omega}^2\cos^2\theta+A_\mu) \big] \\
=& \, 0 \, .
\end{split}
\eeq
This is the local Tayler instability dispersion relation found by \cite{zahn:07}\footnote{We believe there is a typo the dispersion relation of \cite{zahn:07}, originating from a typo in equation (A3) of \cite{spruit:99}. The Alfv\'{e}n velocity should be defined as $\omega_\mathrm{A}=V_\mathrm{A}/\varpi$, as illustrated from equation (2.12) in \cite{acheson:78}. Spruit includes a redundant $m$ in his definition, leading to Zahn's dispersion relation, different from ours by factors of $m$ in a few terms.
\label{foot}
}, but at an arbitrary latitude $\theta$. It differs only by factors of $\cos^2 \theta$, which of course equal unity at the pole, and are of order unity at mid latitudes. 

In Appendix \ref{appendixc2}, we analyze the growing modes of this dispersion relation in the limits of rapid rotation  ($\omega_{\rm A} \ll \Omega$), a magnetic field profile $B_\phi \propto \sin \theta$, and fast thermal diffusion, each of which are likely to be realized in many stars. The key conclusions are
\begin{enumerate}
\item Only $m=1$ modes can grow. Modes with $m=0$ or $m\geq2$ are always damped. 
\item Growing modes exist where $\theta<\pi/3$ or $\theta>2\pi/3$. They do not exist at the equator.
\item Modes only grow in the presence of finite (but not too large) magnetic diffusivity, and the fastest growing modes have growth rates
\beq
\label{taylergamma}
\gamma \sim k_r^2 \eta \sim \frac{\omega_\mathrm{A}^2}{\Omega} \, .
\eeq
\end{enumerate}
These results confirm the findings of \cite{spruit:99,spruit:02,zahn:07} at the pole and generalize them to arbitrary latitudes.

\subsection{Non-linear Saturation}

There is general agreement that equation \ref{taylerdisp} is a linear dispersion relation for Tayler modes, with the fastest growing modes having $m=1$ and growth rate comparable to equation \ref{taylergamma}. However, the non-linear evolution, and likely turbulent saturation of the instability, is not well understood or agreed upon. Similarly, the AM transport via Maxwell stresses in the saturated state remains unclear. \cite{spruit:02} proposed that the instability saturates when a turbulent damping rate is equal to the mode growth rate, and when the magnetic energy generation by winding radial field lines is equal to the magnetic energy dissipation by turbulence. \cite{zahn:07} contested Spruit's argument on the grounds that the dynamo loop proposed to regenerate the radial component of the magnetic field cannot operate. 

\cite{fuller:19} re-examined the non-linear evolution of the Tayler instability. They agreed with \cite{zahn:07} about the dynamo loop closure problem, and used stability considerations to propose an alternative criterion for the strength of the radial component of the magnetic field. They also investigated the rate at which magnetic perturbations non-linearly cascade to smaller scales, equating this with the energy damping rate. \cite{fuller:19} argued that \cite{spruit:02} greatly overestimated the damping rate by assuming that energy in the ordered background field $\vec{B}$ could be damped, when only energy in the disordered (i.e., varying on short length scales) perturbed field $\vec{b}$ cascades to smaller scales to be damped. Whereas \cite{spruit:02} assumed an energy damping rate $\dot{E} \sim \gamma B^2$, with $\gamma$ given by equation \ref{taylergamma}, \cite{fuller:19} estimated an energy damping rate $\dot{E} \sim \gamma |b|^2$.  The smaller energy damping rate of \cite{fuller:19} allows the radial and horizontal magnetic field to grow to larger amplitudes, exherting stronger Maxwell stresses and causing more AM transport than prior predictions.

\subsection{Tayler Torques}
\label{sec2.3}

\cite{fuller:19} estimated the torque density $T$ from Maxwell stresses in the saturated state of the Tayler instability, finding
\beq
T \sim 4 \pi \rho r^2 \alpha^3 q \Omega^2 \bigg(\frac{\Omega}{N_{\rm eff}}\bigg)^2 \, ,
\eeq
where $\alpha$ is a constant of order unity, defined via the saturated Alfv\'{e}n frequency $\omega_{\rm A} = \alpha \Omega (q \Omega/N_{\rm eff})^{1/3}$. We use a fiducial value of $\alpha=1$, which approximately matches asteroseismic measurements of core rotation rates in low-mass stars. The effective stratification is typically $N_{\rm eff} \simeq N_\mu$ except where no signifcant composition gradient exists, as discussed in \cite{fuller:19}. We implement Tayler torques in our models via an effective viscosity operating on gradients of rotation frequency,
\beq
\label{amdiff}
\nu_\mathrm{AM} = r^2 \Omega\bigg(\frac{\Omega}{N_\mathrm{eff}}\bigg)^2 \, .
\eeq
As in \cite{fuller:19}, these toques are only implemented above a critical shear rate 
\beq
\label{qmin}
q_{\rm min} \sim \bigg(\frac{N_{\rm eff}}{\Omega}\bigg)^{5/2} \bigg( \frac{\eta}{r^2 \Omega} \bigg)^{3/4} 
\eeq
because magnetic diffusion eliminates the instability at smaller shears. We assume rotation constant on spheres in radiative regions due to the much more rapid AM transport in the horizontal direction relative to the radial direction. Convective regions of our models are effectively rigidly rotating due to the assumption of a very large effective turbulent viscosity in convective zones.

\section{Results}
\label{sec4}

\subsection{Models}
\label{sec2.4}

We predict internal rotation rates of massive stars using stellar models constructed with the MESA stellar evolution code \citep{paxton:11,paxton:13,paxton:15,paxton:18}. The models include the effective AM diffusivity of equation \ref{amdiff} when the shear exceeds equation \ref{qmin}, while the classic prescription for Tayler torques \citep{spruit:02} in MESA has been turned off, nearly identical to the low-mass models of \cite{fuller:19}. We study massive stars with initial masses ranging from $12-45\,M_\odot$, and roughly solar initial metallicity $Z=0.02$. Our models include mass loss via the ``Dutch" prescription (with efficiency $\eta=0.5$), and moderate convective overshoot. To determine the dependence of remnant rotation rate on progenitor rotation rate, we examine models with initial equatorial rotational velocities of $50,150$ and $450\,\mathrm{km/s}$. Whereas typical massive stars have rotation velocities in the range $50-150$ km/s, 450 km/s represents an extremely rapidly rotating massive star \citep{demink:13}. We run our models from the zero age main sequence (ZAMS) to the moment when the silicon core of a star exceeds $1.5\,M_\odot$, i.e., just before the onset of core silicon burning. At this point, the star is only $\sim$days from core-collapse, and extrapolation of our models predicts no more significant AM transport (see the discussion below).

\begin{figure}
\begin{center}
\includegraphics[scale=0.5]{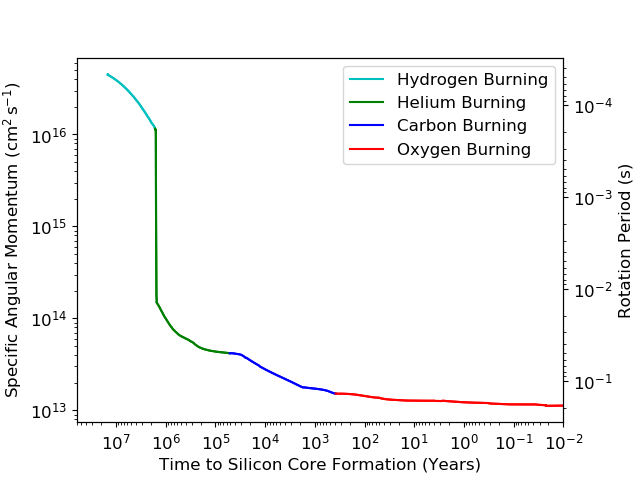}
\end{center} 
\caption{\label{1p5AMT} Specific angular momentum of the inner $1.5\,M_\odot$ core of a star with initial mass $14.0\,M_\odot$ and initial equatorial rotational velocity of $150\,\mathrm{km/s}$. The right axis shows the corresponding neutron star rotation period. The line is color-coded by evolutionary phase: core hydrogen burning (cyan), shell hydrogen and core helium burning (green), shell helium and core carbon burning (blue), and shell carbon/core oxygen burning (red). The vast majority of the core's AM is lost just after the main sequence, as the helium core contracts and the star expands into a red supergiant. The model terminates when the mass of the silicon core exceeds $1.5 \, M_\odot$.
}
\end{figure}

\begin{figure*}
\begin{center}
\includegraphics[scale=0.5]{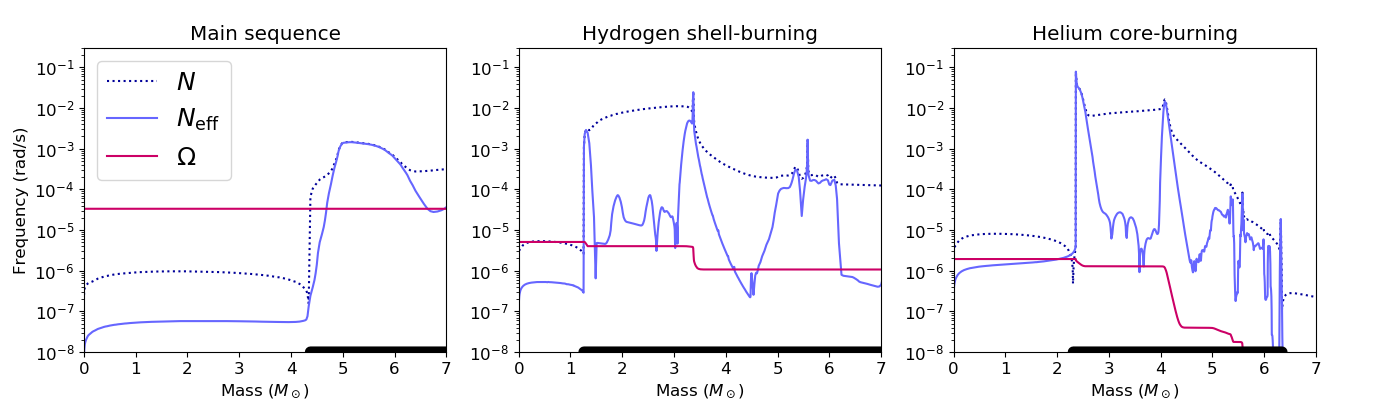}
\end{center} 
\caption{\label{omegac} Profiles of angular rotation frequency $\Omega$ (red line), effective Brunt-V\"ais\"al\"a frequency $N_\mathrm{eff}$ (blue solid line), and Brunt-V\"ais\"al\"a frequency $N$ (blue dashed line) of the same model shown in Figure \ref{1p5AMT}, at three phases of evolution. Radiative zones are labeled with thick black lines on the $x$-axis. {\bf Left:} On the main sequence, the larger ratio of $\Omega/N_{\rm eff}$ allows for efficient angular momentum transport and rigid rotation. {\bf Middle:} During hydrogen shell-burning when the helium core is contracting, core angular momentum is rapidly transported outwards, creating the ``green cliff" in Figure \ref{1p5AMT}. {\bf Right:} During core helium burning, the smaller ratio of $\Omega/N_{\rm eff}$ suppresses Tayler torques, allowing differential rotation to persist, creating the ``green plateau" in Figure \ref{1p5AMT}.
}
\end{figure*}

To gauge the efficiency of AM transport at different phases of evolution, we compute the specific AM contained within a mass coordinate $m$,
\beq
j = \frac{J(m)}{m} \,
\eeq
with 
\beq
J(m) = \int^m_0 \frac{2}{3} r^2 \Omega \, dm \, .
\eeq
In the absence of AM transport, $j$ is constant at every mass coordinate within the star. However, as the core contracts during stellar evolution, its angular velocity increases relative to the envelope, creating shear within the star. Tayler torques act upon this shear, causing the core to spin slower by transporting AM from the core to the envelope. Phases of evolution where the core $j$ decreases the most are most important for determining its pre-collapse rotation rate. A useful diagnostic is the core's specific AM at a mass coordinate $m_{1.5}=1.5 \, M_\odot$, denoted $j(m_{1.5}) = j_{1.5}$, which is a typical baryonic NS mass. In the absence of other effects, the core AM within this mass coordinate determines the natal spin rates of NSs via
\beq
P_{\rm NS} = \frac{2 \pi I_{\rm NS}}{j_{1.5} m_{1.5}} \, ,
\eeq
and we adopt a typical NS moment of inertia $I_{\rm NS} = 10^{45} \, {\rm g} \, {\rm cm}^{2}$. 

\begin{figure}
\begin{center}
\includegraphics[scale=0.5]{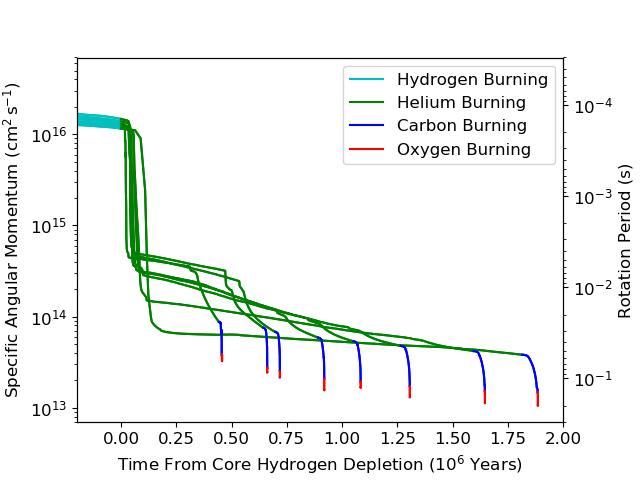}
\end{center} 
\caption{\label{esem1p5} The evolution of specific angular momentum of the $1.5\,M_\odot$ core for our models with initial surface velocity of $150\,\mathrm{km/s}$. As in Figure \ref{1p5AMT}, lines are colored by evolutionary phase, and the right $y$-axis shows the corresponding neutron star rotation period. The different lines correspond to models with different initial masses (end points from left to right, $45\,M_\odot,30\,M_\odot,25\,M_\odot,20\,M_\odot,18\,M_\odot,16\,M_\odot,14\,M_\odot$ and $12\,M_\odot$ ).
}
\end{figure}

Figure \ref{1p5AMT} shows the specific AM $j_{1.5}$ of a $14 \,M_\odot$ stellar model as it evolves, and the corresponding NS rotation rate $P_{\rm NS}$. The different colors correspond to different burning processes during stellar evolution, in the sense that when the central mass fraction of an element has decreased below $10^{-2}$, we define that burning phase to have ceased. Hence, the blue line denotes core hydrogen burning, the green line shows hydrogen shell burning and core helium burning, etc. Note the most significant AM loss occurs immediately after core hydrogen burning, where roughly 99\% of the core's AM is lost. The sudden AM loss occurs as the helium core contracts and the hydrogen envelope expands, generating shear that is damped by Tayler torques. \cite{fuller:19} found very similar behavior in low-mass red giant stars, which lose most of their core AM just after hydrogen depletion, as the helium core contracts and the star moves up the red giant branch.

Figure \ref{omegac} demonstrates why so much AM is lost during hydrogen shell-burning, which is primarily related to the ratio of $\Omega/N_{\rm eff}$ that determines the AM diffusivity of equation \ref{amdiff}. On the main sequence, the value of $\Omega/N_{\rm eff}$ is large enough that AM transport is efficient and nearly rigid rotation is maintained. The core loses a small amount of AM due to the moderate contraction of the core and expansion of the envelope during core hydrogen burning. After the main sequence, the core contracts (and the envelope expands) by a large factor, generating internal shear and allowing Tayler torques to extract large amounts of AM from the core. When the helium core first starts to contract, the value $\Omega/N_{\rm eff} \sim 10^{-3}$ is not too small, allowing most of the core's AM to be extracted. During core helium burning, the value $\Omega/N_{\rm eff} \sim 10^{-4}$ decreases, slowing AM loss from the core and forming the plateau at $\sim 10^5-10^6$ years in Figure \ref{1p5AMT}. Large amounts of differential rotation are present within the star from this stage onward. After core helium burning, the carbon core contracts, again increasing the shear and allowing for more efficient core AM extraction. This produces the steeper fall off of $j_{1.5}$ at $\sim 10^4$ years in Figure \ref{1p5AMT}. Once core carbon burning begins, the core's AM content has nearly reached its final value, and only drops by another factor of $\sim$2 before core-collapse.

In Figure \ref{esem1p5}, we plot the value of $j_{1.5}$ for stars of several initial masses. In each model, the vast majority of the core's AM is lost just after core hydrogen burning during helium core contraction. We see from Figure \ref{esem1p5} that the AM extraction rate during helium core contraction is similar (or even larger) in massive stars. However, in more massive stars, the helium core does not have to contract as much before helium burning begins, such that less AM is extracted by the time the rotation profile reaches the critical shear level (equation \ref{qmin}), after which very little AM is transported. Correspondingly, the more extended helium-burning cores of massive stars contract by a larger factor between the end of helium burning and the beginning of oxygen burning. In these higher mass stars, more AM is extracted during carbon/oxygen core contraction, compared to that extracted during helium core contraction. The two effects somewhat balance, such that the values of $j_{1.5}$ in massive stars at the start of oxygen burning are only a few times larger than in lower mass stars. While the AM extraction rate during core carbon/oxygen burning is larger than it is during helium burning (i.e., the slope $d \log j_{1.5}/dt$ in Figure \ref{esem1p5} is large during carbon/oxygen burning), the duration of these phases is very short. Hence, the fraction of core AM lost during carbon/oxygen burning is much smaller than that lost during hydrogen shell-burning.

\begin{figure}
\begin{center}
\includegraphics[scale=0.5]{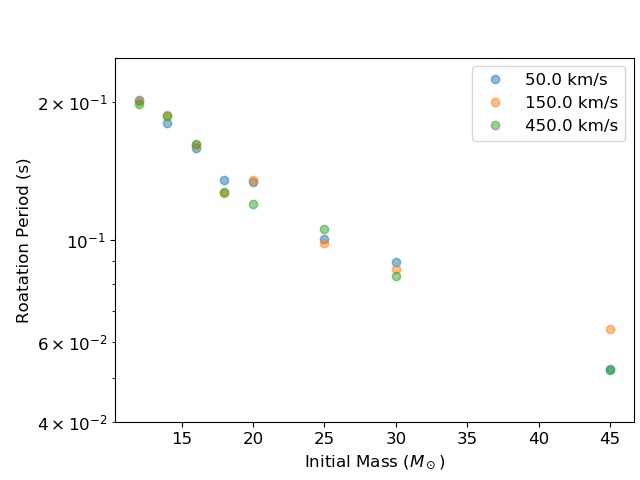}
\end{center} 
\caption{\label{NST} Predicted natal neutron star rotation rates, as a function of progenitor mass, assuming conservation of core angular momentum during silicon burning and core-collapse. The blue, orange and green points correspond to initial rotational velocities of $50, 150$ and $450\,\mathrm{km/s}$, respectively. The orange points correspond to the endpoints of the tracks shown in Figure \ref{esem1p5}. 
}
\end{figure}

\begin{figure}
\begin{center}
\includegraphics[scale=0.6]{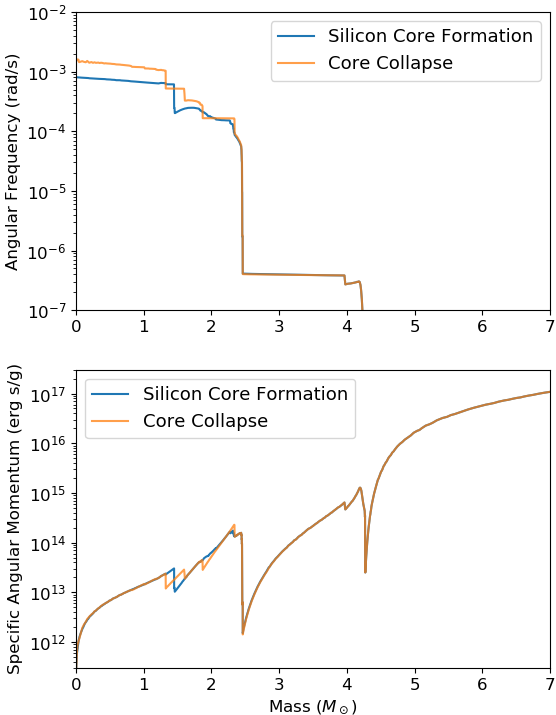}
\end{center} 
\caption{\label{Fprofile} Angular frequency (top panel) and specific angular momentum (bottom panel) of the same model shown in Figure \ref{1p5AMT} near oxygen depletion (silicon core formation) and near core-collapse.
}
\end{figure}

The value of $j_{1.5}$ at the end of each track in Figures \ref{1p5AMT} and \ref{esem1p5} corresponds to the expected natal NS rotation period $P_{\rm NS}$ shown by the right axis. Figure \ref{NST} shows these values of $P_{\rm NS}$ for each of our models, with different initial surface velocities marked by different colors. These models predict decreasing NS rotation periods with increasing stellar mass, but all of our models have $P_{\rm NS}$ in the range $50-200\,\mathrm{ms}$. As discussed above, the faster NS spins of high-mass models stems from the fact that their cores have a larger physical extent that allows them to retain more AM.

Remarkably, for most models, the final NS rotation periods have little dependence on the initial surface velocity or total AM of the star. This result is similar to that of \cite{fuller:19} for low-mass stars, who found post-main sequence internal rotation rates scaled very weakly with initial rotation rate, due to the convergent evolution in core rotation rate caused by the strong scaling of equation \ref{amdiff} with rotation rate, and the fact that AM tends to be transported out of the core until the minimum shear of equation \ref{qmin} is realized. The dispersion of final rotation periods for stars of the same mass in Figure \ref{NST} appears to stem from slightly different evolutionary histories (due to small numerical differences coupled with ``chaotic'' stellar evolution as described by \citealt{sukhbold:18}), rather than being directly tied to the initial rotation rate. We find that models with higher resolution produce nearly identical results (suggesting our models are largely converged) apart from some small residual scatter similar to that shown in Figure \ref{NST}.

Finally, we must check whether Tayler torques can operate as prescribed by equation \ref{amdiff} during all phases of evolution of our models. The instability can no longer reach its equilibrium and transport AM according to equation \ref{amdiff} if the Tayler mode growth time $t_T \sim \Omega/\omega_{\rm A}^2$ is longer than the AM transport timescale $t_{\rm AM} \sim r^2/\nu_{\rm AM} \sim N_{\rm eff}^2/\Omega^3$. Using the saturated Alfv\'en frequency from \cite{fuller:19}, $\omega_{\rm A} \sim \Omega (q \Omega / N_{\rm eff})^{1/3}$, estimating $N_{\rm eff} \sim \sqrt{G \rho}$, $j = \Omega r^2$, $\rho \sim m r^{-3}$ and using $q \sim 1$, we find $t_{\rm T}/t_{\rm AM} > 1$ when $r_{\rm min} \lesssim j^2/(G m)$. Evaluating this timescale at a mass coordinate $m=1.4 \, M_\odot$ at the end of our runs yields $r_{\rm min} \sim 1 \, {\rm cm}$, which is much smaller than the corresponding radial coordinate. Hence our models are always in an acceptable regime. We note that our models can reach a regime where $t_{\rm T}$ is longer than the evolution time, which can occur around the silicon burning phase. However, the AM transport time $t_{\rm AM}$ at this phase is much longer than the remaining lifetime of the star, so the effects of Tayler torques are negligible. 

The pre-collapse AM distribution may affect the dynamics of the SN explosion process \citep{foglizzo:17}. Figure \ref{Fprofile} shows profiles of the angular frequency and specific AM within the 14 $M_\odot$ model shown in Figure \ref{1p5AMT}. We have plotted these quantities at two times, when the silicon core and the iron core reach $\approx 1.4 \, M_\odot$. We find a pre-collapse rotation frequency of $\Omega \sim 2 \times 10^{-3}$ rad/s in the iron core, corresponding to a specific AM of $j \sim 10^{13} \, {\rm cm}^2 \, {\rm s}^{-1}$. Sharp gradients in the rotation frequency are associated with composition gradients between different burning shells. Neither the rotation profile nor the AM profile are flat in the core. At the end of the star's evolution, Tayler torques become inefficient at redistributing AM within radiative zones, but convective torques still enforce profiles with constant $\Omega$ and hence $j \propto r^2$. After convection subsides, shells change their radial coordinate and their rotation rate, but their specific AM is approximately conserved. The final rotation profile thus depends on the details of the star's evolution, but the AM profile as a function of mass remains roughly constant.

\section{Discussion}
\label{sec5}

The AM transport predicted by our models has important implications for massive star evolution. During the main sequence, we find the core and envelope remain tightly coupled, with little internal shear. Hence, prior stellar evolution calculations exhibiting mixing due to shear instabilities (e.g., \citealt{meynet:97,georgy:13}) have likely overestimated the extent of mixing, though rotational mixing via Eddington-Sweet circulation can still occur. This mixing is typically most important on the main sequence, where both our model and the Tayler-Spruit dynamo \citep{spruit:02} predict nearly rigid rotation. Main sequence rotational mixing for such models are unlikely to be greatly altered, but post-main sequence mixing should be re-examined in light of the slower internal rotation rates we predict. 

Our calculations have crucial implications for the supernova explosions that mark their deaths, and for the spins of compact object remnants left behind. For our simple models of single stars at solar metallicity, the prediction is clear: AM transport is efficient, massive stellar cores rotate slowly, and compact objects are likely to be born slowly rotating.

For most NSs, our models predict initial spin periods $P_{\rm NS} \sim 100-200 \, {\rm ms}$ as shown in Figure \ref{NST}, nearly independent of initial rotation velocity of the star. We caution that these predictions have some dependence on the Tayler instability saturation parameter $\alpha$, which \cite{fuller:19} found to be $\alpha \approx 1$ based on calibration by asteroseismic measurements of low-mass stars. There is also dependence on the uncertain moment of inertia $I$ for typical NSs, but neither effect produces uncertaintly larger than a factor of $\sim \! 2$. However, red giant cores and white dwarfs exhibit scatter by a factor of a few around the predicted rotation periods. It would not be surprising if a similar scatter exists for massive star cores, potentially producing NSs with natal rotation periods of tens to hundreds of milliseconds.

Interestingly, our predictions are near the initial spin rates estimated for typical pulsars \citep{faucher:06,popov:10,popov:12,gullon:14}, which likely have a broad distribution of tens to hundreds of milliseconds. Our  predictions are also similar to the spins of central compact object (CCO) pulsars, which have $P_{\rm NS} \sim 100-400 \, {\rm ms}$ \citep{halpern:10,gotthelf:13}. Since the spin-down times of these ``anti-magnetar'' COOs are orders of magnitude larger than their ages, the observed spin period is nearly identical to their natal spin period, justifying a direct comparison with our predictions.

However, it is clear that some pulsars are born spinning faster than our estimates, e.g., the Crab pulsar which is estimated to have had a natal spin rate of $P_{\rm NS} \sim 20 \, {\rm ms}$ \citep{kaspi:02}. We posit that most NSs have had their spin rates altered by stochastic AM transport processes during the final stages of massive star evolution or during the SN explosion \citep{spruit:98}. \cite{fullerwave:15} considered the AM transported by internal gravity waves excited by the vigorous convection due to late nuclear burning stages. The stochastic influx of AM into the core of the star results in a Maxwellian distribution of NS spin frequencies with typical natal spin periods of $P_{\rm NS} \sim 50-100 \, {\rm ms}$. Additionally, three-dimensional simulations of core-collapse SNe demonstrate asymmetric explosions that stochastically spin the NS to periods as short as $P_{\rm NS} \sim 20 \, {\rm ms}$ \citep{muller:18}, though prompt explosions of low-mass or ultra-stripped progenitors produce much less spin-up \citep{muller:18b,rantsiou:11}. Given the slow core rotation predicted by our models, these stochastic mechanisms could be the dominant processes that determine natal NS rotation periods. In these scenarios, the rotation rate and spin-axis of a young NS is essentially uncorrelated with the spin of its progenitor star. Indeed, the spin-orbit misalignment of the more recently formed NS in the binary pulsar system PSR J0737-3039 is consistent with this hypothesis \citep{farr:11}. The relatively slow spins of CCOs may indicate they originated from prompt explosions, with spins inherited from their progenitor star.

Very rapidly rotating NSs and BHs are almost certainly formed under rare circumstances, and such objects are thought to be the central engines of broad-lined Ic SNe, gamma-ray bursts, and superluminous SNe (e.g., \citealt{woosley:93,kasen:10,metzger:11}). We believe that close binary evolution or homogeneous evolution (e.g., \citealt{yoon:06}) may be required to form these rapidly rotating compact objects. Tidal spin-up of a helium star progenitor (e.g., \citealt{kushnir:16,qin:18}) will allow for rapidly rotating BHs. Whether this scneario can allow for rapidly rotating NSs is unclear, as Tayler torques may be able to couple the inner core of the star with the overlying helium envelope, preventing rapid core rotation. Preliminary binary runs indicate this to be the case, but we defer a more detailed treatment of binary scenarios to future work. 

The leading alternative to our models is that of \cite{kissin:15,kissin:18}. In stark contrast to our models, those models feature rigid rotation in radiative regions (enforced by fossil fields) and differential rotation in convective zones (resulting from inward convective AM pumping). Interestingly, these models also predict NS rotation periods of $\sim$hundreds of milliseconds for typical progenitors. But they allow for much more rapid NS rotation for certain stars (e.g., $25 \, M_\odot$) with very thick surface/core convective zones, and very slow NS rotation for more massive stars that have lost most of their AM to winds. Distinguishing between our model and theirs will benefit from more detailed comparisons with data for low-mass stars.

\section{Conclusion}
\label{sec6}

We have examined angular momentum (AM) transport in massive stars via the baroclinic and Tayler instabilities. In contrast to many previous works \citep{knobloch:82,knobloch:83,spruit:84,zahn:92}, and in agreement with other works \citep{tassoul:82,fujimoto:87,kitchatinov:14}, we find that unstable baroclinic modes generally exist in differentially rotating radiative zones, when thermal diffusion is neglected. In most stars, however, these modes will be stabilized by thermal diffusion, such that baroclinic instabilities are unlikely to play an important role in AM redistribution. 

We also analyze the Tayler instability at arbitrary latitudes inside of stars, finding that unstable modes likely do not exist near the stellar equator and growth rates peak near the pole of the star. Typical growth rates and wavenumbers are similar to those found in prior work (e.g., \citealt{spruit:99,spruit:02,zahn:07}). We then apply the Tayler instability saturation criterion and AM transport prescription of \cite{fuller:19} to massive stellar models using MESA. We find that after the main sequence, roughly 99\% of core AM is lost during hydrogen shell-burning while the helium core is contracting and the star is expanding into a red supergiant. The combination of large shear and modest buoyancy frequency during this stage allows for efficient AM transport, preventing the spin-up of the contracting core as its AM is transferred to the expanding envelope. A smaller, but still significant, amount of AM is lost during helium burning and beyond. The small core spin rates and internal shears of our models imply that rotational mixing effects may be somewhat overestimated in stellar models not taking magnetic torques into account.

Assuming AM conservation during core-collapse and the subsequent supernova explosion, our models predict natal rotation periods of $P\sim 100-200$ ms for NSs born from progenitors of $12-20 \, M_\odot$. We predict faster rotation for higher-mass stars, but even our $45 \, M_\odot$ model would produce a relatively slowly rotating NS with $P\approx 50$ ms. Hence, our models predict rotation periods similar to those inferred for newborn pulsars \citep{faucher:06,popov:10,popov:12,gullon:14}, though perhaps near the slow end of the distribution. In light of the slow core rotation of our models, the spins of many NSs may be primarily determined by stochastic AM redistribution processes just before core-collapse \citep{fullerwave:15} or during the explosion process \citep{spruit:98}. Relatively rapidly rotating pulsars like the Crab pulsar ($P_i \sim 20 \, {\rm ms}$, \citealt{kaspi:02}) can be accounted for by core-collapse spin-up as found in some supernova simulations \citep{muller:17,muller:18}.

Because AM is generally transported from the contracting helium core into the expanding hydrogen envelope, our models predict that many black holes will be born slowly rotating, which we investigate in a companion paper (Fuller \& Ma, in prep). However, it is important to emphasize that the majority of massive stars evolve in interacting binaries \citep{sana:12}, and hence the predictions of our single-star models are merely one piece of the puzzle. We suspect AM accretion via mass transfer and tidal spin-up in close binaries will be able to produce more rapidly rotating compact objects capable of powering various energetic transients, a subject we plan to investigate in future work.

\section{Acknowledgements}
We thank Yulin Gong for helpful comments on the manuscript, and Adam Jermyn and Anthony Piro for useful insights on the baroclinic instability. This research is funded in part by an Innovator Grant from The Rose Hills Foundation, the Sloan Foundation through grant FG-2018-10515, and by the Gordon and Betty Moore Foundation through Grant GBMF7392.

\bibliography{CoreRotBib}

\begin{thebibliography}{}
\makeatletter
\relax
\def\mn@urlcharsother{\let\do\@makeother \do\$\do\&\do\#\do\^\do\_\do\%\do\~}
\def\mn@doi{\begingroup\mn@urlcharsother \@ifnextchar [ {\mn@doi@}
  {\mn@doi@[]}}
\def\mn@doi@[#1]#2{\def\@tempa{#1}\ifx\@tempa\@empty \href
  {http://dx.doi.org/#2} {doi:#2}\else \href {http://dx.doi.org/#2} {#1}\fi
  \endgroup}
\def\mn@eprint#1#2{\mn@eprint@#1:#2::\@nil}
\def\mn@eprint@arXiv#1{\href {http://arxiv.org/abs/#1} {{\tt arXiv:#1}}}
\def\mn@eprint@dblp#1{\href {http://dblp.uni-trier.de/rec/bibtex/#1.xml}
  {dblp:#1}}
\def\mn@eprint@#1:#2:#3:#4\@nil{\def\@tempa {#1}\def\@tempb {#2}\def\@tempc
  {#3}\ifx \@tempc \@empty \let \@tempc \@tempb \let \@tempb \@tempa \fi \ifx
  \@tempb \@empty \def\@tempb {arXiv}\fi \@ifundefined
  {mn@eprint@\@tempb}{\@tempb:\@tempc}{\expandafter \expandafter \csname
  mn@eprint@\@tempb\endcsname \expandafter{\@tempc}}}

\bibitem[\protect\citeauthoryear{{Abbott} et~al.,}{{Abbott}
  et~al.}{2016}]{ligorun1}
{Abbott} B.~P.,  et~al., 2016, \mn@doi [Physical Review X]
  {10.1103/PhysRevX.6.041015}, \href
  {http://adsabs.harvard.edu/abs/2016PhRvX...6d1015A} {6, 041015}

\bibitem[\protect\citeauthoryear{{Abbott} et~al.,}{{Abbott}
  et~al.}{2017a}]{ligo170104}
{Abbott} B.~P.,  et~al., 2017a, \mn@doi [Physical Review Letters]
  {10.1103/PhysRevLett.118.221101}, \href
  {http://adsabs.harvard.edu/abs/2017PhRvL.118v1101A} {118, 221101}

\bibitem[\protect\citeauthoryear{{Abbott} et~al.,}{{Abbott}
  et~al.}{2017b}]{ligo170814}
{Abbott} B.~P.,  et~al., 2017b, \mn@doi [Physical Review Letters]
  {10.1103/PhysRevLett.119.141101}, \href
  {http://adsabs.harvard.edu/abs/2017PhRvL.119n1101A} {119, 141101}

\bibitem[\protect\citeauthoryear{{Abbott} et~al.,}{{Abbott}
  et~al.}{2017c}]{ligo170608}
{Abbott} B.~P.,  et~al., 2017c, \mn@doi [\apjl] {10.3847/2041-8213/aa9f0c},
  \href {http://adsabs.harvard.edu/abs/2017ApJ...851L..35A} {851, L35}

\bibitem[\protect\citeauthoryear{{Acheson} \& {Gibbons}}{{Acheson} \&
  {Gibbons}}{1978}]{acheson:78}
{Acheson} D.~J.,  {Gibbons} M.~P.,  1978, \mn@doi [Philosophical Transactions
  of the Royal Society of London Series A] {10.1098/rsta.1978.0066}, \href
  {http://adsabs.harvard.edu/abs/1978RSPTA.289..459A} {289, 459}

\bibitem[\protect\citeauthoryear{{Beck} et~al.,}{{Beck} et~al.}{2012}]{beck:12}
{Beck} P.~G.,  et~al., 2012, \mn@doi [\nat] {10.1038/nature10612}, \href
  {http://adsabs.harvard.edu/abs/2012Natur.481...55B} {481, 55}

\bibitem[\protect\citeauthoryear{{Benomar}, {Takata}, {Shibahashi}, {Ceillier}
  \& {Garc{\'{\i}}a}}{{Benomar} et~al.}{2015}]{benomar:15}
{Benomar} O.,  {Takata} M.,  {Shibahashi} H.,  {Ceillier} T.,   {Garc{\'{\i}}a}
  R.~A.,  2015, \mn@doi [\mnras] {10.1093/mnras/stv1493}, \href
  {http://adsabs.harvard.edu/abs/2015MNRAS.452.2654B} {452, 2654}

\bibitem[\protect\citeauthoryear{{Bildsten}, {Ushomirsky}  \&
  {Cutler}}{{Bildsten} et~al.}{1996}]{bildsten:96}
{Bildsten} L.,  {Ushomirsky} G.,   {Cutler} C.,  1996, \mn@doi [\apj]
  {10.1086/177012}, \href {http://adsabs.harvard.edu/abs/1996ApJ...460..827B}
  {460, 827}

\bibitem[\protect\citeauthoryear{{Cantiello}, {Mankovich}, {Bildsten},
  {Christensen-Dalsgaard}  \& {Paxton}}{{Cantiello}
  et~al.}{2014}]{cantiello:14}
{Cantiello} M.,  {Mankovich} C.,  {Bildsten} L.,  {Christensen-Dalsgaard} J.,
  {Paxton} B.,  2014, \mn@doi [\apj] {10.1088/0004-637X/788/1/93}, \href
  {http://adsabs.harvard.edu/abs/2014ApJ...788...93C} {788, 93}

\bibitem[\protect\citeauthoryear{{Charney} \& {Stern}}{{Charney} \&
  {Stern}}{1962}]{charney:62}
{Charney} J.~G.,  {Stern} M.~E.,  1962, \mn@doi [Journal of Atmospheric
  Sciences] {10.1175/1520-0469(1962)019<0159:OTSOIB>2.0.CO;2}, \href
  {http://adsabs.harvard.edu/abs/1962JAtS...19..159C} {19, 159}

\bibitem[\protect\citeauthoryear{{Deheuvels} et~al.,}{{Deheuvels}
  et~al.}{2014}]{deheuvels:14}
{Deheuvels} S.,  et~al., 2014, \mn@doi [\aap] {10.1051/0004-6361/201322779},
  \href {http://adsabs.harvard.edu/abs/2014A%26A...564A..27D} {564, A27}

\bibitem[\protect\citeauthoryear{{Deheuvels}, {Ballot}, {Beck}, {Mosser},
  {{\O}stensen}, {Garc{\'{\i}}a}  \& {Goupil}}{{Deheuvels}
  et~al.}{2015}]{deheuvels:15}
{Deheuvels} S.,  {Ballot} J.,  {Beck} P.~G.,  {Mosser} B.,  {{\O}stensen} R.,
  {Garc{\'{\i}}a} R.~A.,   {Goupil} M.~J.,  2015, \mn@doi [\aap]
  {10.1051/0004-6361/201526449}, \href
  {http://adsabs.harvard.edu/abs/2015A%26A...580A..96D} {580, A96}

\bibitem[\protect\citeauthoryear{{Denissenkov} \& {Pinsonneault}}{{Denissenkov}
  \& {Pinsonneault}}{2007}]{denissenkov:07}
{Denissenkov} P.~A.,  {Pinsonneault} M.,  2007, \mn@doi [\apj]
  {10.1086/510345}, \href {http://adsabs.harvard.edu/abs/2007ApJ...655.1157D}
  {655, 1157}

\bibitem[\protect\citeauthoryear{{Farr}, {Kremer}, {Lyutikov}  \&
  {Kalogera}}{{Farr} et~al.}{2011}]{farr:11}
{Farr} W.~M.,  {Kremer} K.,  {Lyutikov} M.,   {Kalogera} V.,  2011, \mn@doi
  [\apj] {10.1088/0004-637X/742/2/81}, \href
  {http://adsabs.harvard.edu/abs/2011ApJ...742...81F} {742, 81}

\bibitem[\protect\citeauthoryear{{Faucher-Gigu{\`e}re} \&
  {Kaspi}}{{Faucher-Gigu{\`e}re} \& {Kaspi}}{2006}]{faucher:06}
{Faucher-Gigu{\`e}re} C.-A.,  {Kaspi} V.~M.,  2006, \mn@doi [\apj]
  {10.1086/501516}, \href {http://adsabs.harvard.edu/abs/2006ApJ...643..332F}
  {643, 332}

\bibitem[\protect\citeauthoryear{{Foglizzo}}{{Foglizzo}}{2017}]{foglizzo:17}
{Foglizzo} T.,  2017, {Explosion Physics of Core-Collapse Supernovae}.
p.~1053, \mn@doi{10.1007/978-3-319-21846-5_52}

\bibitem[\protect\citeauthoryear{{Fujimoto}}{{Fujimoto}}{1987}]{fujimoto:87}
{Fujimoto} M.~Y.,  1987, \aap, \href
  {http://adsabs.harvard.edu/abs/1987A%26A...176...53F} {176, 53}

\bibitem[\protect\citeauthoryear{{Fujimoto}}{{Fujimoto}}{1988}]{fujimoto:88}
{Fujimoto} M.~Y.,  1988, \aap, \href
  {http://adsabs.harvard.edu/abs/1988A%26A...198..163F} {198, 163}

\bibitem[\protect\citeauthoryear{{Fuller}, {Cantiello}, {Lecoanet}  \&
  {Quataert}}{{Fuller} et~al.}{2015}]{fullerwave:15}
{Fuller} J.,  {Cantiello} M.,  {Lecoanet} D.,   {Quataert} E.,  2015, \mn@doi
  [\apj] {10.1088/0004-637X/810/2/101}, \href
  {http://adsabs.harvard.edu/abs/2015ApJ...810..101F} {810, 101}

\bibitem[\protect\citeauthoryear{{Fuller}, {Piro}  \& {Jermyn}}{{Fuller}
  et~al.}{2019}]{fuller:19}
{Fuller} J.,  {Piro} A.~L.,   {Jermyn} A.~S.,  2019, \mn@doi [\mnras]
  {10.1093/mnras/stz514}, \href
  {https://ui.adsabs.harvard.edu/abs/2019MNRAS.485.3661F} {485, 3661}

\bibitem[\protect\citeauthoryear{{Gehan}, {Mosser}, {Michel}, {Samadi}  \&
  {Kallinger}}{{Gehan} et~al.}{2018}]{gehan:18}
{Gehan} C.,  {Mosser} B.,  {Michel} E.,  {Samadi} R.,   {Kallinger} T.,  2018,
  \mn@doi [\aap] {10.1051/0004-6361/201832822}, \href
  {http://adsabs.harvard.edu/abs/2018A%26A...616A..24G} {616, A24}

\bibitem[\protect\citeauthoryear{{Georgy}, {Ekstr{\"o}m}, {Granada}, {Meynet},
  {Mowlavi}, {Eggenberger}  \& {Maeder}}{{Georgy} et~al.}{2013}]{georgy:13}
{Georgy} C.,  {Ekstr{\"o}m} S.,  {Granada} A.,  {Meynet} G.,  {Mowlavi} N.,
  {Eggenberger} P.,   {Maeder} A.,  2013, \mn@doi [\aap]
  {10.1051/0004-6361/201220558}, \href
  {http://adsabs.harvard.edu/abs/2013A%26A...553A..24G} {553, A24}

\bibitem[\protect\citeauthoryear{{Gotthelf}, {Halpern}  \& {Alford}}{{Gotthelf}
  et~al.}{2013}]{gotthelf:13}
{Gotthelf} E.~V.,  {Halpern} J.~P.,   {Alford} J.,  2013, \mn@doi [\apj]
  {10.1088/0004-637X/765/1/58}, \href
  {http://adsabs.harvard.edu/abs/2013ApJ...765...58G} {765, 58}

\bibitem[\protect\citeauthoryear{{Gull{\'o}n}, {Miralles}, {Vigan{\`o}}  \&
  {Pons}}{{Gull{\'o}n} et~al.}{2014}]{gullon:14}
{Gull{\'o}n} M.,  {Miralles} J.~A.,  {Vigan{\`o}} D.,   {Pons} J.~A.,  2014,
  \mn@doi [\mnras] {10.1093/mnras/stu1253}, \href
  {http://adsabs.harvard.edu/abs/2014MNRAS.443.1891G} {443, 1891}

\bibitem[\protect\citeauthoryear{{Halpern} \& {Gotthelf}}{{Halpern} \&
  {Gotthelf}}{2010}]{halpern:10}
{Halpern} J.~P.,  {Gotthelf} E.~V.,  2010, \mn@doi [\apj]
  {10.1088/0004-637X/709/1/436}, \href
  {http://adsabs.harvard.edu/abs/2010ApJ...709..436H} {709, 436}

\bibitem[\protect\citeauthoryear{{Heger}, {Langer}  \& {Woosley}}{{Heger}
  et~al.}{2000}]{heger:00}
{Heger} A.,  {Langer} N.,   {Woosley} S.~E.,  2000, \mn@doi [\apj]
  {10.1086/308158}, \href {http://adsabs.harvard.edu/abs/2000ApJ...528..368H}
  {528, 368}

\bibitem[\protect\citeauthoryear{{Heger}, {Woosley}  \& {Spruit}}{{Heger}
  et~al.}{2005}]{heger:05}
{Heger} A.,  {Woosley} S.~E.,   {Spruit} H.~C.,  2005, \mn@doi [\apj]
  {10.1086/429868}, \href {http://adsabs.harvard.edu/abs/2005ApJ...626..350H}
  {626, 350}

\bibitem[\protect\citeauthoryear{{Hermes} et~al.,}{{Hermes}
  et~al.}{2017}]{hermes:17}
{Hermes} J.~J.,  et~al., 2017, \mn@doi [\apjs] {10.3847/1538-4365/aa8bb5},
  \href {http://adsabs.harvard.edu/abs/2017ApJS..232...23H} {232, 23}

\bibitem[\protect\citeauthoryear{Kasen \& Bildsten}{Kasen \&
  Bildsten}{2010}]{kasen:10}
Kasen D.,  Bildsten L.,  2010, \mn@doi [{ApJ}] {10.1088/0004-637x/717/1/245},
  717, 245

\bibitem[\protect\citeauthoryear{{Kasen} \& {Woosley}}{{Kasen} \&
  {Woosley}}{2009}]{kasen:09}
{Kasen} D.,  {Woosley} S.~E.,  2009, \mn@doi [\apj]
  {10.1088/0004-637X/703/2/2205}, \href
  {http://adsabs.harvard.edu/abs/2009ApJ...703.2205K} {703, 2205}

\bibitem[\protect\citeauthoryear{{Kaspi} \& {Helfand}}{{Kaspi} \&
  {Helfand}}{2002}]{kaspi:02}
{Kaspi} V.~M.,  {Helfand} D.~J.,  2002, in {Slane} P.~O.,  {Gaensler} B.~M.,
  eds,  Astronomical Society of the Pacific Conference Series Vol. 271, Neutron
  Stars in Supernova Remnants. p.~3 (\mn@eprint {} {astro-ph/0201183})

\bibitem[\protect\citeauthoryear{{Kissin} \& {Thompson}}{{Kissin} \&
  {Thompson}}{2015}]{kissin:15}
{Kissin} Y.,  {Thompson} C.,  2015, \mn@doi [\apj]
  {10.1088/0004-637X/808/1/35}, \href
  {http://adsabs.harvard.edu/abs/2015ApJ...808...35K} {808, 35}

\bibitem[\protect\citeauthoryear{{Kissin} \& {Thompson}}{{Kissin} \&
  {Thompson}}{2018}]{kissin:18}
{Kissin} Y.,  {Thompson} C.,  2018, \mn@doi [\apj] {10.3847/1538-4357/aab1fb},
  \href {http://adsabs.harvard.edu/abs/2018ApJ...862..111K} {862, 111}

\bibitem[\protect\citeauthoryear{{Kitchatinov}}{{Kitchatinov}}{2014}]{kitchatinov:14}
{Kitchatinov} L.~L.,  2014, \mn@doi [\apj] {10.1088/0004-637X/784/1/81}, \href
  {http://adsabs.harvard.edu/abs/2014ApJ...784...81K} {784, 81}

\bibitem[\protect\citeauthoryear{{Knobloch} \& {Spruit}}{{Knobloch} \&
  {Spruit}}{1982}]{knobloch:82}
{Knobloch} E.,  {Spruit} H.~C.,  1982, \aap, \href
  {http://adsabs.harvard.edu/abs/1982A%26A...113..261K} {113, 261}

\bibitem[\protect\citeauthoryear{{Knobloch} \& {Spruit}}{{Knobloch} \&
  {Spruit}}{1983}]{knobloch:83}
{Knobloch} E.,  {Spruit} H.~C.,  1983, \aap, \href
  {http://adsabs.harvard.edu/abs/1983A%26A...125...59K} {125, 59}

\bibitem[\protect\citeauthoryear{{Kushnir}, {Zaldarriaga}, {Kollmeier}  \&
  {Waldman}}{{Kushnir} et~al.}{2016}]{kushnir:16}
{Kushnir} D.,  {Zaldarriaga} M.,  {Kollmeier} J.~A.,   {Waldman} R.,  2016,
  \mn@doi [\mnras] {10.1093/mnras/stw1684}, \href
  {http://adsabs.harvard.edu/abs/2016MNRAS.462..844K} {462, 844}

\bibitem[\protect\citeauthoryear{{Lee} \& {Saio}}{{Lee} \&
  {Saio}}{1997}]{lee:97}
{Lee} U.,  {Saio} H.,  1997, \apj, \href
  {http://adsabs.harvard.edu/abs/1997ApJ...491..839L} {491, 839}

\bibitem[\protect\citeauthoryear{{MacFadyen} \& {Woosley}}{{MacFadyen} \&
  {Woosley}}{1999}]{macfadyen:99}
{MacFadyen} A.~I.,  {Woosley} S.~E.,  1999, \mn@doi [\apj] {10.1086/307790},
  \href {http://adsabs.harvard.edu/abs/1999ApJ...524..262M} {524, 262}

\bibitem[\protect\citeauthoryear{Maeder \& Meynet}{Maeder \&
  Meynet}{2000}]{Maeder_2000}
Maeder A.,  Meynet G.,  2000, \mn@doi [Annual Review of Astronomy and
  Astrophysics] {10.1146/annurev.astro.38.1.143}, 38, 143

\bibitem[\protect\citeauthoryear{Metzger, Giannios, Thompson, Bucciantini  \&
  Quataert}{Metzger et~al.}{2011}]{metzger:11}
Metzger B.~D.,  Giannios D.,  Thompson T.~A.,  Bucciantini N.,   Quataert E.,
  2011, \mn@doi [Monthly Notices of the Royal Astronomical Society]
  {10.1111/j.1365-2966.2011.18280.x}, 413, 2031

\bibitem[\protect\citeauthoryear{{Meynet} \& {Maeder}}{{Meynet} \&
  {Maeder}}{1997}]{meynet:97}
{Meynet} G.,  {Maeder} A.,  1997, \aap, \href
  {http://adsabs.harvard.edu/abs/1997A%26A...321..465M} {321, 465}

\bibitem[\protect\citeauthoryear{{Miller} \& {Miller}}{{Miller} \&
  {Miller}}{2015}]{miller:15}
{Miller} M.~C.,  {Miller} J.~M.,  2015, \mn@doi [\physrep]
  {10.1016/j.physrep.2014.09.003}, \href
  {http://adsabs.harvard.edu/abs/2015PhR...548....1M} {548, 1}

\bibitem[\protect\citeauthoryear{{Mosser} et~al.,}{{Mosser}
  et~al.}{2012}]{mosser:12}
{Mosser} B.,  et~al., 2012, \mn@doi [\aap] {10.1051/0004-6361/201220106}, \href
  {http://adsabs.harvard.edu/abs/2012A%26A...548A..10M} {548, A10}

\bibitem[\protect\citeauthoryear{{M{\"u}ller}, {Melson}, {Heger}  \&
  {Janka}}{{M{\"u}ller} et~al.}{2017}]{muller:17}
{M{\"u}ller} B.,  {Melson} T.,  {Heger} A.,   {Janka} H.-T.,  2017, \mn@doi
  [\mnras] {10.1093/mnras/stx1962}, \href
  {http://adsabs.harvard.edu/abs/2017MNRAS.472..491M} {472, 491}

\bibitem[\protect\citeauthoryear{{M{\"u}ller} et~al.,}{{M{\"u}ller}
  et~al.}{2018a}]{muller:18}
{M{\"u}ller} B.,  et~al., 2018a, preprint, \href
  {http://adsabs.harvard.edu/abs/2018arXiv181105483M} {} (\mn@eprint {arXiv}
  {1811.05483})

\bibitem[\protect\citeauthoryear{{M{\"u}ller}, {Gay}, {Heger}, {Tauris}  \&
  {Sim}}{{M{\"u}ller} et~al.}{2018b}]{muller:18b}
{M{\"u}ller} B.,  {Gay} D.~W.,  {Heger} A.,  {Tauris} T.~M.,   {Sim} S.~A.,
  2018b, \mn@doi [\mnras] {10.1093/mnras/sty1683}, \href
  {http://adsabs.harvard.edu/abs/2018MNRAS.479.3675M} {479, 3675}

\bibitem[\protect\citeauthoryear{{Nicholl}, {Guillochon}  \&
  {Berger}}{{Nicholl} et~al.}{2017}]{nicholl:17}
{Nicholl} M.,  {Guillochon} J.,   {Berger} E.,  2017, \mn@doi [\apj]
  {10.3847/1538-4357/aa9334}, \href
  {http://adsabs.harvard.edu/abs/2017ApJ...850...55N} {850, 55}

\bibitem[\protect\citeauthoryear{{Paxton}, {Bildsten}, {Dotter}, {Herwig},
  {Lesaffre}  \& {Timmes}}{{Paxton} et~al.}{2011}]{paxton:11}
{Paxton} B.,  {Bildsten} L.,  {Dotter} A.,  {Herwig} F.,  {Lesaffre} P.,
  {Timmes} F.,  2011, \mn@doi [\apjs] {10.1088/0067-0049/192/1/3}, \href
  {http://adsabs.harvard.edu/abs/2011ApJS..192....3P} {192, 3}

\bibitem[\protect\citeauthoryear{{Paxton} et~al.,}{{Paxton}
  et~al.}{2013}]{paxton:13}
{Paxton} B.,  et~al., 2013, \mn@doi [\apjs] {10.1088/0067-0049/208/1/4}, \href
  {http://adsabs.harvard.edu/abs/2013ApJS..208....4P} {208, 4}

\bibitem[\protect\citeauthoryear{{Paxton} et~al.,}{{Paxton}
  et~al.}{2015}]{paxton:15}
{Paxton} B.,  et~al., 2015, \mn@doi [\apjs] {10.1088/0067-0049/220/1/15}, \href
  {http://adsabs.harvard.edu/abs/2015ApJS..220...15P} {220, 15}

\bibitem[\protect\citeauthoryear{{Paxton} et~al.,}{{Paxton}
  et~al.}{2018}]{paxton:18}
{Paxton} B.,  et~al., 2018, \mn@doi [\apjs] {10.3847/1538-4365/aaa5a8}, \href
  {http://adsabs.harvard.edu/abs/2018ApJS..234...34P} {234, 34}

\bibitem[\protect\citeauthoryear{Pedlosky}{Pedlosky}{1992}]{pedlosky:92}
Pedlosky J.,  1992, {Geophysical Fluid Dynamics}.
Springer study edition, Springer New York, \url
  {https://books.google.com/books?id=FXs-uRSDBFYC}

\bibitem[\protect\citeauthoryear{{Piro} \& {Bildsten}}{{Piro} \&
  {Bildsten}}{2007}]{piro:07}
{Piro} A.~L.,  {Bildsten} L.,  2007, \mn@doi [\apj] {10.1086/518687}, \href
  {http://adsabs.harvard.edu/abs/2007ApJ...663.1252P} {663, 1252}

\bibitem[\protect\citeauthoryear{{Popov} \& {Turolla}}{{Popov} \&
  {Turolla}}{2012}]{popov:12}
{Popov} S.~B.,  {Turolla} R.,  2012, \mn@doi [\apss]
  {10.1007/s10509-012-1100-z}, \href
  {http://adsabs.harvard.edu/abs/2012Ap%26SS.341..457P} {341, 457}

\bibitem[\protect\citeauthoryear{{Popov}, {Pons}, {Miralles}, {Boldin}  \&
  {Posselt}}{{Popov} et~al.}{2010}]{popov:10}
{Popov} S.~B.,  {Pons} J.~A.,  {Miralles} J.~A.,  {Boldin} P.~A.,   {Posselt}
  B.,  2010, \mn@doi [\mnras] {10.1111/j.1365-2966.2009.15850.x}, \href
  {http://adsabs.harvard.edu/abs/2010MNRAS.401.2675P} {401, 2675}

\bibitem[\protect\citeauthoryear{{Qin}, {Marchant}, {Fragos}, {Meynet}  \&
  {Kalogera}}{{Qin} et~al.}{2018a}]{qin:18b}
{Qin} Y.,  {Marchant} P.,  {Fragos} T.,  {Meynet} G.,   {Kalogera} V.,  2018a,
  preprint, \href {http://adsabs.harvard.edu/abs/2018arXiv181013016Q} {}
  (\mn@eprint {arXiv} {1810.13016})

\bibitem[\protect\citeauthoryear{{Qin}, {Fragos}, {Meynet}, {Andrews},
  {S{\o}rensen}  \& {Song}}{{Qin} et~al.}{2018b}]{qin:18}
{Qin} Y.,  {Fragos} T.,  {Meynet} G.,  {Andrews} J.,  {S{\o}rensen} M.,
  {Song} H.~F.,  2018b, \mn@doi [\aap] {10.1051/0004-6361/201832839}, \href
  {http://adsabs.harvard.edu/abs/2018A%26A...616A..28Q} {616, A28}

\bibitem[\protect\citeauthoryear{{Rantsiou}, {Burrows}, {Nordhaus}  \&
  {Almgren}}{{Rantsiou} et~al.}{2011}]{rantsiou:11}
{Rantsiou} E.,  {Burrows} A.,  {Nordhaus} J.,   {Almgren} A.,  2011, \mn@doi
  [\apj] {10.1088/0004-637X/732/1/57}, \href
  {https://ui.adsabs.harvard.edu/abs/2011ApJ...732...57R} {732, 57}

\bibitem[\protect\citeauthoryear{{Sana} et~al.,}{{Sana} et~al.}{2012}]{sana:12}
{Sana} H.,  et~al., 2012, \mn@doi [Science] {10.1126/science.1223344}, \href
  {http://adsabs.harvard.edu/abs/2012Sci...337..444S} {337, 444}

\bibitem[\protect\citeauthoryear{{Spruit}}{{Spruit}}{1999}]{spruit:99}
{Spruit} H.~C.,  1999, \aap, \href
  {http://adsabs.harvard.edu/abs/1999A%26A...349..189S} {349, 189}

\bibitem[\protect\citeauthoryear{{Spruit}}{{Spruit}}{2002}]{spruit:02}
{Spruit} H.~C.,  2002, \mn@doi [\aap] {10.1051/0004-6361:20011465}, \href
  {http://adsabs.harvard.edu/abs/2002A%26A...381..923S} {381, 923}

\bibitem[\protect\citeauthoryear{{Spruit} \& {Knobloch}}{{Spruit} \&
  {Knobloch}}{1984}]{spruit:84}
{Spruit} H.~C.,  {Knobloch} E.,  1984, \aap, \href
  {http://adsabs.harvard.edu/abs/1984A%26A...132...89S} {132, 89}

\bibitem[\protect\citeauthoryear{{Spruit} \& {Phinney}}{{Spruit} \&
  {Phinney}}{1998}]{spruit:98}
{Spruit} H.,  {Phinney} E.~S.,  1998, \mn@doi [\nat] {10.1038/30168}, \href
  {http://adsabs.harvard.edu/abs/1998Natur.393..139S} {393, 139}

\bibitem[\protect\citeauthoryear{{Spruit}, {Knobloch}  \& {Roxburgh}}{{Spruit}
  et~al.}{1983}]{spruit:83}
{Spruit} H.~C.,  {Knobloch} E.,   {Roxburgh} I.~W.,  1983, \mn@doi [\nat]
  {10.1038/304520a0}, \href {http://adsabs.harvard.edu/abs/1983Natur.304..520S}
  {304, 520}

\bibitem[\protect\citeauthoryear{{Sukhbold}, {Woosley}  \& {Heger}}{{Sukhbold}
  et~al.}{2018}]{sukhbold:18}
{Sukhbold} T.,  {Woosley} S.~E.,   {Heger} A.,  2018, \mn@doi [\apj]
  {10.3847/1538-4357/aac2da}, \href
  {http://adsabs.harvard.edu/abs/2018ApJ...860...93S} {860, 93}

\bibitem[\protect\citeauthoryear{{Tassoul} \& {Tassoul}}{{Tassoul} \&
  {Tassoul}}{1982}]{tassoul:82}
{Tassoul} J.-L.,  {Tassoul} M.,  1982, \mn@doi [\apjs] {10.1086/190801}, \href
  {http://adsabs.harvard.edu/abs/1982ApJS...49..317T} {49, 317}

\bibitem[\protect\citeauthoryear{{Tayler}}{{Tayler}}{1973}]{tayler:73}
{Tayler} R.~J.,  1973, \mn@doi [\mnras] {10.1093/mnras/161.4.365}, \href
  {http://adsabs.harvard.edu/abs/1973MNRAS.161..365T} {161, 365}

\bibitem[\protect\citeauthoryear{{The LIGO Scientific Collaboration}
  et~al.,}{{The LIGO Scientific Collaboration} et~al.}{2018}]{ligoo2b:18}
{The LIGO Scientific Collaboration} et~al., 2018, arXiv e-prints, \href
  {http://adsabs.harvard.edu/abs/2018arXiv181112940T} {}

\bibitem[\protect\citeauthoryear{{Valsecchi}, {Glebbeek}, {Farr}, {Fragos},
  {Willems}, {Orosz}, {Liu}  \& {Kalogera}}{{Valsecchi}
  et~al.}{2010}]{valsecchi:10}
{Valsecchi} F.,  {Glebbeek} E.,  {Farr} W.~M.,  {Fragos} T.,  {Willems} B.,
  {Orosz} J.~A.,  {Liu} J.,   {Kalogera} V.,  2010, \mn@doi [\nat]
  {10.1038/nature09463}, \href
  {http://adsabs.harvard.edu/abs/2010Natur.468...77V} {468, 77}

\bibitem[\protect\citeauthoryear{{Wheeler}, {Kagan}  \&
  {Chatzopoulos}}{{Wheeler} et~al.}{2015}]{wheeler:15}
{Wheeler} J.~C.,  {Kagan} D.,   {Chatzopoulos} E.,  2015, \mn@doi [\apj]
  {10.1088/0004-637X/799/1/85}, \href
  {http://adsabs.harvard.edu/abs/2015ApJ...799...85W} {799, 85}

\bibitem[\protect\citeauthoryear{{Woosley}}{{Woosley}}{1993}]{woosley:93}
{Woosley} S.~E.,  1993, \mn@doi [\apj] {10.1086/172359}, \href
  {http://adsabs.harvard.edu/abs/1993ApJ...405..273W} {405, 273}

\bibitem[\protect\citeauthoryear{Yoon, Langer  \& Norman}{Yoon
  et~al.}{2006}]{yoon:06}
Yoon S.-C.,  Langer N.,   Norman C.,  2006, \mn@doi [Astronomy and
  Astrophysics] {10.1051/0004-6361:20065912}, 460, 199

\bibitem[\protect\citeauthoryear{{Zahn}}{{Zahn}}{1992}]{zahn:92}
{Zahn} J.-P.,  1992, \aap, \href
  {http://adsabs.harvard.edu/abs/1992A%26A...265..115Z} {265, 115}

\bibitem[\protect\citeauthoryear{{Zahn} \& {Zinn-Justin}}{{Zahn} \&
  {Zinn-Justin}}{1993}]{zahn:93}
{Zahn} J.-P.,  {Zinn-Justin} J.,  eds, 1993, {Astrophysical Fluid Dynamics. Les
  Houches Session LXVII}

\bibitem[\protect\citeauthoryear{{Zahn}, {Brun}  \& {Mathis}}{{Zahn}
  et~al.}{2007}]{zahn:07}
{Zahn} J.-P.,  {Brun} A.~S.,   {Mathis} S.,  2007, \mn@doi [\aap]
  {10.1051/0004-6361:20077653}, \href
  {http://adsabs.harvard.edu/abs/2007A%26A...474..145Z} {474, 145}

\bibitem[\protect\citeauthoryear{{Zaldarriaga}, {Kushnir}  \&
  {Kollmeier}}{{Zaldarriaga} et~al.}{2018}]{zaldarriaga:18}
{Zaldarriaga} M.,  {Kushnir} D.,   {Kollmeier} J.~A.,  2018, \mn@doi [\mnras]
  {10.1093/mnras/stx2577}, \href
  {http://adsabs.harvard.edu/abs/2018MNRAS.473.4174Z} {473, 4174}

\bibitem[\protect\citeauthoryear{{de Mink}, {Langer}, {Izzard}, {Sana}  \& {de
  Koter}}{{de Mink} et~al.}{2013}]{demink:13}
{de Mink} S.~E.,  {Langer} N.,  {Izzard} R.~G.,  {Sana} H.,   {de Koter} A.,
  2013, \mn@doi [\apj] {10.1088/0004-637X/764/2/166}, \href
  {http://adsabs.harvard.edu/abs/2013ApJ...764..166D} {764, 166}

\makeatother
\end{thebibliography}

\appendix

\begin{table}
    \centering
    \begin{tabular}{l|r}
      $\vec{b}$ & Perturbed magnetic field\\
      $\vec{B}$ & Magnetic field\\
      $B_0$ & Background magnetic field magnitude\\
      $c_P$ & Specific heat at constant pressure\\
      $c_\text{s}$ & Adiabatic sound speed\\
      $c_T$ & Isothermal sound speed\\
      $c_V$ & Specific heat at constant volume\\
      $\hat{e}$ & Unit vector in $\vec{e}$ direction\\
      $f$ & Coriolis frequency\\
      $F$ & Angular overlap integral\\
      $g$ & Gravitational acceleration\\
      $G$ & Gravitational constant\\
      $H$ & Pressure scale height\\
      $H_\lambda$ & Hough function\\
      $I_\text{NS}$ & Neutron star moment of inertia\\
      $j$ & Specific angular momentum\\
      $J$ & Angular momentum\\
      $k$ & Total wave number \\
      $(k_r,l,m)$ & Wave number in $(r,\theta,\phi)$ direction\\
      $(k_x,k_y,k_z)$ & Wave number in $(x,y,z)$ direction\\
      $k_\perp$ & Wave number in horizontal direction\\
      $\vec{L}$ & Perturbed Lorentz force\\
      $\ell$ & Wave length\\
      $\ell_r$ & Wave length in radial direction\\
      $\ell_h$ & Wave length in horizontal direction\\
      $m$ & Mass\\
      $m_\text{p}$ & Proton mass\\
      $M$ & Total mass\\
      $M_\odot$ & Solar mass\\
      $N$ & Brunt-V\"ais\"al\"a frequency\\
      $N_\text{eff}$ & Effective Brunt-V\"ais\"al\"a frequency\\
      $N_T$ & Thermal part of Brunt-V\"ais\"al\"a frequency\\
      $N_\theta$ & Baroclinic characteristic frequency\\
      $N_\mu$ & Composition part of Brunt-V\"ais\"al\"a frequency\\
      $P$ & Pressure\\
      $P_\text{NS}$ & Rotation period of neutron star\\
      $q$ & Dimensionless shear\\
      $(r,\theta,\phi)$ & Spherical coordinates \\
      $r_c$ & Baroclinic critical radius\\
      $S$ & Entropy\\
      $t$ & Time\\
      $T$ & Temperature\\
      $\vec{v}$ & Velocity\\
      $(x,y,z)$ & Local Cartesian coordinates \\
      $\alpha$ & Torque parameter\\
      $\gamma$ & Growth rate\\
      $\Gamma_1$ & Adiabatic index\\
      $\delta Q$ & Eulerian perturbation of $Q$\\
      $\eta$ & Magnetic diffusivity\\
      $\kappa$ & Thermal diffusivity\\
      $\lambda$ & Eigenvalue of Hough function\\
      $\mu$ & Composition\\
      $\vec{\xi}$ & Lagrangian displacement vector\\
      $\rho$ & Density\\
      $\nu$ & Rotation parameter\\
      $\nu_\text{AM}$ & Effective viscosity\\
      $\omega$ & Angular oscillation frequency\\
      $\omega_A$ & Alfv\'{e}n frequency\\
      $\omega_\text{c}$ & Critical Alfv\'{e}n frequency\\
      $\Omega\,(\vec{\Omega})$ & Angular rotation frequency (vector)\\
    \end{tabular}
    \caption{List of variables used in main text and appendices}
    \label{tab:my_label}
\end{table}

\section{Energy Equation with Thermal Diffusion and Baroclinic Term}
\label{appendixa}

Here we derive the energy equation for an ideal gas with thermal diffusion and baroclinicity. We work in a corotating frame such that the background velocity $v$ is zero, and we assume an axisymmetric background structure. The entropy for an ideal gas is
\beq
S=c_V\ln P-c_P\ln \rho \, 
\eeq
and the Brunt-V\"ais\"al\"a frequency is
\beq
\label{eq2}
N^2=\frac{g}{\Gamma_1}\frac{\partial \ln P}{\partial r}-g\frac{\partial \ln\rho}{\partial r}=\frac{g}{c_P}\frac{\partial S}{\partial r} \,
\eeq
where $\Gamma_1=c_P/c_V$. We assume that temperature diffusion is the only process that changes entropy, which is a good approximation for short wavelength disturbances deep inside a star, such that
\begin{align}
dS&=\frac{c_PdT}{T}\\
\Rightarrow\;\frac{dS}{dt}&=\frac{c_P}{T}\frac{dT}{dt}=\frac{c_P\kappa}{T}\nabla^2T\\
\Rightarrow\;\frac{\partial S}{\partial t}&+\vec{v}\cdot\nabla S=\frac{c_P\kappa}{T}\nabla^2T \, .
\end{align}
Taking the Eulerian perturbation and assuming time dependence of each perturbed variable $\delta Q$,
\beq
\delta Q\propto e^{-i\omega t} \,
\eeq
we have
\begin{align}
\label{dsdt}
\frac{\partial \delta S}{\partial t}+\delta \vec{v}\cdot\nabla S&=\frac{c_P\kappa}{T}\nabla^2\delta T\\
\Rightarrow\;-i\omega(\delta S +\vec{\xi}\cdot\nabla S)&=-\frac{c_P\kappa}{T}k^2\delta T\\
\Rightarrow\;-i\omega\bigg(\delta S +\xi_r \frac{\partial S}{\partial r}+&\xi_\theta \frac{1}{r}\frac{\partial S}{\partial \theta}\bigg)=-\frac{c_P\kappa}{T}k^2\delta T \, .
\end{align}
To obtain equation \ref{dsdt} we have used a standard WKB approximation by setting $\nabla^2 \delta T = - k^2 \delta T$, where $k$ is the wavenumber of the perturbation, and dropping terms on the right hand side with weaker $k$ dependence. 

Next, following \cite{kitchatinov:14}, we neglect higher order centrifugal terms (valid in relatively slowly rotating stars) by defining our radial coordinate $r$ to be perpendicular to isobars $r$ so that
\beq
\label{dsdtheta}
\xi_\theta\frac{1}{r}\frac{\partial S}{\partial \theta}=\xi_\theta\frac{1}{r}\bigg(-\frac{c_P}{\rho}\frac{\partial \rho}{\partial \theta}\bigg)=\xi_\theta\frac{c_P}{g}\bigg(-\frac{g}{\rho r}\frac{\partial \rho}{\partial \theta}\bigg) \, .
\eeq
Next, we assume rotation constant on spherical shells, due to efficient horizontal AM transport within radiative regions of a star. Defining the latitudinal Brunt-V\"ais\"al\"a frequency 
\beq
\label{eq12}
N_\theta^2 \equiv -\frac{g}{\rho r}\frac{\partial \rho}{\partial \theta} \, ,
\eeq
and using the baroclinicity relation for shellular rotation (e.g., \citealt{kitchatinov:14}), we have
\beq
\label{eq13}
N_\theta^2 = 2q\Omega^2\sin\theta\cos\theta \, ,
\eeq
where 
\beq
q=d\ln\Omega/d\ln r
\eeq
is the dimensionless shear. The latitudinal Brunt-V\"ais\"al\"a frequency $N_\theta^2$ is a horizontal buoyancy force due to baroclinicity arrising from differential rotation.

Combining equations \ref{dsdtheta}, \ref{eq12}, and \ref{eq2}, we have
\beq
\label{eq14}
-i\omega\bigg(\delta S+\xi_r\frac{c_P}{g}N^2+\xi_\theta\frac{c_P}{g}N_\theta^2\bigg)=-\frac{c_P\kappa}{T}k^2\delta T \, .
\eeq
For an ideal gas
\begin{align}
P&=\frac{\rho k_\mathrm{B} T}{\mu m_\mathrm{p}}\\
\Rightarrow\;\frac{\delta P}{P}=&\frac{\delta \rho}{\rho}+\frac{\delta T}{T}-\frac{\delta \mu}{\mu} \, .
\end{align}
The Lagrangian perturbation $\Delta \mu$ is related to the Eulerian perturbation $\delta \mu$ by
\begin{align}
\Delta\mu=\delta\mu&+\vec{\xi}\cdot\nabla\mu=0\\
\Rightarrow\;\delta\mu&=-\xi_r\frac{\partial\mu}{\partial r} \,
\end{align}
where we have assumed $\mu$ has only $r$ dependence, again due to efficient latitudinal transport of composition and AM. Then
\beq
\frac{\delta T}{T}=\frac{\delta P}{P}-\frac{\delta \rho}{\rho}-\xi_r\frac{\partial \ln \mu}{\partial r}
\eeq
together with \eqref{eq14}, we have
\beq
\begin{split}
i\omega\bigg(\delta S+\xi_r\frac{c_P}{g}N^2&+\xi_\theta\frac{c_P}{g}N_\theta^2\bigg)\\
=c_P\kappa k^2\bigg(\frac{\delta P}{P}&-\frac{\delta \rho}{\rho}-\xi_r\frac{\partial \ln \mu}{\partial r}\bigg) \, .
\end{split}
\eeq
Assuming $c_P$ and $c_V$ to be constant (as they are for an ideal gas), we have
\beq
\delta S=c_P\bigg(\frac{1}{\Gamma_1}\frac{\delta P}{P}-\frac{\delta \rho}{\rho}\bigg) \, .
\eeq
Defining the compositional buoyancy frequency
\beq
N_\mu^2\equiv-g\partial \ln\mu/\partial r \, ,
\eeq
we have
\beq
\begin{split}
i\omega\bigg(\frac{1}{\Gamma_1}\frac{\delta P}{P}-\frac{\delta \rho}{\rho}+\xi_r\frac{N^2}{g}&+\xi_\theta\frac{N_\theta^2}{g}\bigg)\\
=\kappa k^2\bigg(\frac{\delta P}{P}&-\frac{\delta \rho}{\rho}+\xi_r\frac{N^2_\mu}{g}\bigg)
\end{split}
\eeq
\beq
\begin{split}
\Rightarrow\;\bigg(1+\frac{i\kappa k^2}{\omega}\bigg)\frac{\delta \rho}{\rho}&=\bigg(\frac{1}{\Gamma_1}+\frac{i\kappa k^2}{\omega}\bigg)\frac{\rho}{P}\frac{\delta P}{\rho}\\
&+\xi_r\frac{N^2}{g}+\xi_\theta\frac{N_\theta^2}{g}+\xi_r\frac{N^2_\mu}{g}\frac{i\kappa k^2}{\omega} \, .
\end{split}
\eeq
Note that
\beq
N^2=N_T^2+N_\mu^2 \,
\eeq
where $N_T$ is the thermal component of the buoyancy frequency. Then
\beq
\begin{split}
\bigg(1+\frac{i\kappa k^2}{\omega}\bigg)\frac{\delta \rho}{\rho}&=\bigg(\frac{1}{\Gamma_1}+\frac{i\kappa k^2}{\omega}\bigg)\frac{\rho}{P}\frac{\delta P}{\rho}\\
+\xi_r\frac{N_T^2}{g}&+\xi_\theta\frac{N_\theta^2}{g}+\xi_r\frac{N^2_\mu}{g}\bigg(1+\frac{i\kappa k^2}{\omega}\bigg) \, ,
\end{split}
\eeq
can be rewritten
\beq
\label{EE}
\frac{\delta \rho}{\rho}=\frac{1}{\tilde{c_\mathrm{s}}^2}\frac{\delta P}{\rho}+\xi_r\frac{\tilde{N}^2}{g}+\xi_\theta\frac{\tilde{N_\theta}^2}{g} \, ,
\eeq
where
\beq
\begin{split}
\frac{1}{\tilde{c_\mathrm{s}}^2}&=\bigg(\frac{1}{\Gamma_1}+\frac{i\kappa k^2}{\omega}\bigg)\bigg(1+\frac{i\kappa k^2}{\omega}\bigg)^{-1}\frac{\rho}{P} \, ,\\
\tilde{N}^2&=\tilde{N_T}^2+N_\mu^2=N_T^2\bigg(1+\frac{i\kappa k^2}{\omega}\bigg)^{-1}+N_\mu^2 \, , \\
\tilde{N_\theta}^2&=N_\theta^2\bigg(1+\frac{i\kappa k^2}{\omega}\bigg)^{-1} \, .
\end{split}
\eeq

Equation \ref{EE} is the energy equation with thermal diffusion. When thermal diffusion is not important, i.e. $\kappa\rightarrow0$,
\beq
\begin{split}
\tilde{c_\mathrm{s}}^2&\sim\frac{\Gamma_1 P}{\rho}=c_s^2 \, , \\
\tilde{N}^2&\sim N_T^2+N_\mu^2=N^2 \, , \\
\tilde{N_\theta}^2&\sim N_\theta^2 \, ,
\end{split}
\eeq
where $c_\text{s}$ is the adiabatic sound speed. The energy equation reduces to
\beq
\frac{\delta \rho}{\rho}=\frac{1}{c_\text{s}^2}\frac{\delta P}{\rho}+\xi_r\frac{N^2}{g}+\xi_\theta\frac{N_\theta^2}{g} \, ,
\eeq
which is the adiabatic energy equation with a baroclinic term. When thermal diffusion is strong, i.e. $\kappa\rightarrow\infty$
\beq
\begin{split}
\tilde{c_\mathrm{s}}^2&\sim\frac{P}{\rho}=c_T^2 \, \\
\tilde{N}^2&\sim N_\mu^2 \, \\
\tilde{N_\theta}^2&\sim0 \,
\end{split}
\eeq
where $c_T$ is the isothermal sound speed. In this limit, the energy equation reduces to
\beq
\label{energydiff}
\frac{\delta \rho}{\rho}=\frac{1}{c_T^2}\frac{\delta P}{\rho}+\xi_r\frac{N_\mu^2}{g} \, .
\eeq
The thermal and baroclinic terms are suppressed by thermal diffusion, and only the composition gradient is important. We  note that the baroclinic term has vanished due to our assumption of composition being constant on isobars. If a latitudinal composition gradient can persist, equation \ref{energydiff} would contain a compositional baroclinic term.

\section{Growth Rate of the Baroclinic Instability}
\label{appendixb}

Because baroclinicity and Coriolis forces break the spherical symmetry of a star, calculating the frequencies and growth rates of global scale modes becomes more difficult. Here we derive the growth rate of the baroclinic instability using the traditional approximation and the geostrophic approximation, following the assumptions made in the main text.

\subsection{Traditional Approximation}
\label{appendixb1}

We start by defining perturbation variables to have time and spatial dependence
\beq
\delta Q\propto \delta Q(\theta)\exp \bigg[ i\bigg(\int k_r dr+m\phi-\omega t\bigg) \bigg] \, .
\eeq
The equation of motion (equation \eqref{eqM}) can be expressed in spherical coordinates
\beq
\label{eq35}
\rho\omega^2\xi_r=\frac{\partial \delta P}{\partial r}+g\delta \rho+2i\Omega\omega\rho\xi_\phi\sin\theta \, ,
\eeq
\beq
\label{eq36}
\rho\omega^2\xi_\theta = \frac{1}{r}\frac{\partial\delta P}{\partial\theta} + 2i\Omega\omega\rho\xi_\phi\cos\theta \, ,
\eeq
\beq
\label{eq37}
\rho\omega^2\xi_\phi = \frac{im}{r\sin\theta}\delta P - 2i\Omega\omega\rho(\xi_\theta\cos\theta+\xi_r\sin\theta) \, .
\eeq
Since the Coriolis force in the radial direction is expected to be much smaller than the buoyancy force, equation \eqref{eq35} reduces to
\beq
\label{eq38}
\rho\omega^2\xi_r=\frac{\partial \delta P}{\partial r}+g\delta\rho \, .
\eeq
We further expect the radial perturbation to be small, such that $\xi_r\ll\xi_\theta$, hence equation \eqref{eq37} reduces to
\beq
\label{eq39}
-\rho\omega^2\xi_\phi=-\frac{im}{r\sin\theta}\delta P+2i\Omega\omega\rho\xi_\theta\cos\theta \, .
\eeq
These approximations are known as the traditional approximation \citep[see e.g.,][]{bildsten:96,lee:97}. It is straightforward to solve for $\xi_\theta$ and $\xi_\phi$ in terms of $\delta P$ from the above equations, yielding
\begin{align}
\xi_\theta&=\frac{1}{\rho\omega^2}\frac{\nu\mu\frac{m}{r\sin\theta}+\frac{1}{r}\frac{\partial}{\partial\theta}}{1-\nu^2\mu^2}\delta P \, \\
\xi_\phi&=\frac{i}{\rho\omega^2}\frac{\frac{m}{r\sin\theta}+\nu\mu\frac{1}{r}\frac{\partial}{\partial\theta}}{1-\nu^2\mu^2}\delta P \,
\end{align}
where we have defined $\nu=2\Omega/\omega$ and $\mu=\cos\theta$. The angular parts of the equation of motion then yield the relation
\beq
\label{xithetaphi}
\frac{1}{\sin\theta}\frac{\partial}{\partial\theta}(\xi_\theta\sin\theta)+\frac{1}{\sin\theta}\frac{\partial\xi_\phi}{\partial\phi}=\frac{1}{\rho\omega^2 r}\hat{\mathcal{L}_\mu}\delta P \,
\eeq
where
\beq
\begin{split}
\hat{\mathcal{L}_\mu}\equiv&\frac{\partial}{\partial\mu}\bigg(\frac{1-\mu^2}{1-\nu^2\mu^2}\frac{\partial}{\partial \mu}\bigg)\\
&-\frac{m^2}{(1-\mu^2)(1-\nu^2\mu^2)}-\frac{m\nu(1+\nu^2\mu^2)}{(1-\nu^2\mu^2)^2} \, .
\end{split}
\eeq

Now consider the continuity equation
\beq
\label{cont}
\begin{split}
&\delta \rho+\nabla\cdot(\rho\vec{\xi})\\
=&\delta \rho+\nabla\rho\cdot\vec{\xi}+\rho\nabla\cdot\vec{\xi}\\
=&\delta\rho+\frac{\partial\rho}{\partial r}\xi_r+\frac{1}{r}\frac{\partial\rho}{\partial\theta}\xi_\theta+\rho\nabla\cdot\vec{\xi}=0
\end{split}
\eeq
We use the incompressible approximation to neglect the first two terms in equation \ref{cont}, but we keep the $\xi_\theta$ term in order to preserve its baroclinic effect. Then we have 
\beq
\begin{split}
\frac{1}{r}\frac{\partial\rho}{\partial\theta}\xi_\theta&+\rho\nabla\cdot\vec{\xi}=\frac{1}{r}\frac{\partial\rho}{\partial\theta}\xi_\theta+\rho\bigg(\frac{1}{r^2}\frac{\partial(r^2\xi_r)}{\partial r}\\
+&\frac{1}{r\sin\theta}\frac{\partial}{\partial\theta}(\xi_\theta\sin\theta)+\frac{1}{r\sin\theta}\frac{\partial\xi_\phi}{\partial\phi}\bigg)=0 \, .
\end{split}
\eeq
Inserting equation \ref{xithetaphi} and making a WKB approximation, we find 
\beq
\frac{1}{r}\frac{\partial \rho}{\partial \theta}\xi_\theta+\rho\bigg(ik_r\xi_r+\frac{1}{\rho\omega^2 r^2}\hat{\mathcal{L}_\mu}\delta P\bigg)=0 \, .
\eeq
Recalling the definition of $N_\theta^2$ in equation \ref{eq12}, and substituting the expression for $\xi_\theta$, we have
\beq
\label{eq49}
\bigg( \hat{\mathcal{L}_\mu} - r\frac{N_\theta^2}{g}\hat{\mathcal{L}_y} \bigg) \delta P=-i\rho\omega^2r^2k_r\xi_r \, ,
\eeq
where
\beq
\hat{\mathcal{L}_y}\equiv \frac{1}{1-\nu^2\mu^2}\bigg(\nu\mu\frac{m}{\sin\theta}+\frac{\partial}{\partial \theta}\bigg) \, .
\eeq
From the definition of $\hat{\mathcal{L}_\mu}$ and $\hat{\mathcal{L}_y}$, we have $\hat{\mathcal{L}_\mu}\sim\hat{\mathcal{L}_y}\sim 1$ for the modes with smallest angular wavenumber. For all but the fastest rotating stars, $N_\theta^2r/g\sim\Omega^2r/g\ll1$, thus we can neglect the second term in the left-hand side of equation \ref{eq49} to find
\beq
\label{eq51}
\hat{\mathcal{L}_\mu}\delta P\approx-i\rho\omega^2r^2k_r\xi_r \, .
\eeq

Now, combining the energy equation \eqref{EE} with the radial equation of motion \eqref{eq38}, we have
\beq
\rho\omega^2\xi_r=\frac{\partial\delta P}{\partial r}+\frac{g}{\tilde{c_\mathrm{s}}^2}\delta P+\rho\tilde{N}^2\xi_r+\rho\tilde{N_\theta}^2\xi_\theta \, .
\eeq
We then make the WKB approximation $\partial\delta P/\partial r\approx ik_r\delta P \approx i k \delta P \gg g\delta P/\tilde{c_\mathrm{s}}^2$. Expressing $\xi_\theta$ in terms of $\delta P$, and limiting ourselves to low-frequency oscillations with $\tilde{N}^2\gg\omega^2$, we find
\beq
-\rho\tilde{N}^2\xi_r=ik_r\delta P+\frac{\tilde{N_\theta}^2}{\omega^2r}\hat{\mathcal{L}_y}\delta P \, .
\eeq
Now combining with equation \ref{eq51}, we cancel the variable $\xi_r$ to obtain 
\beq
\label{eq58}
\bigg(\hat{\mathcal{L}_\mu}-i\frac{\tilde{N_\theta}^2}{\tilde{N}^2}k_rr\hat{\mathcal{L}_y}\bigg)\delta P=-\frac{\omega^2}{\tilde{N}^2}r^2k_r^2\delta P \, .
\eeq

For very low-frequency modes ($\omega \ll \Omega \ll N$) obeying the traditional approximation, $k_r \approx (\tilde{N}/\omega)k_\perp$, and $\hat{\mathcal{L}_y} \sim r k_\perp \omega/\Omega$, so the baroclinic term in equation \ref{eq58} is of order $(q \Omega/\tilde{N}) r^2 k_\perp^2$, while the first term is of order $r^2 k_\perp^2$. For higher frequency modes with $\omega > \Omega$, the baroclinic term is smaller by a factor $\Omega/\omega$. Hence we can treat the baroclinicity as a small perturbation. The unperturbed equation is
\beq
\hat{\mathcal{L}_\mu}\delta P=-\frac{\omega^2}{\tilde{N}^2}r^2k_r^2\delta P
\eeq
The eigenfunctions of the operator $\hat{\mathcal{L}_\mu}$ are Hough functions $H_\lambda(\mu)$ satisfying
\beq
\hat{\mathcal{L}_\mu}H_\lambda(\mu)=-\lambda H_\lambda(\mu) \, .
\eeq
To first order in the baroclinic perturbation, we have the dispersion relation
\beq
\lambda + i\frac{\tilde{N_\theta}^2}{\tilde{N}^2}k_rrF= \frac{\omega^2}{\tilde{N}^2}r^2k_r^2 \, ,
\eeq
where
\beq
F\equiv-\int^1_{-1} H^*_\lambda(\mu)\hat{\mathcal{L}_y}H_\lambda(\mu)d\mu
\eeq
is a dimensionless number depending on $\nu$. We thus obtain the dispersion relation
\beq
\label{eqB28}
\omega^2=\frac{\tilde{N}^2\lambda}{k_r^2r^2}+\frac{i}{k_rr} F \tilde{N_\theta}^2 \, ,
\eeq
which is identical to the usual dispersion relation for Hough modes, but modified to include thermal diffusion and with a baroclinic driving term proportional to $N_\theta^2$.

To solve for the growth rate of baroclinic modes, we first consider the case with small thermal diffusivity, so to zeroth order (when thermal diffusion is neglected), $\tilde{N}^2\sim N^2$, $\tilde{N}_\theta^2\sim N_\theta^2$. Again using the fact that the baroclinic term is much smaller than the buoyancy term, we find 
\beq
\label{eqB29}
\begin{split}
\omega &\simeq \pm \frac{\tilde{N}\lambda^\frac{1}{2}}{|r k_r|} \bigg(1+ir k_r\frac{N_\theta^2}{\tilde{N}^2}\frac{F}{\lambda}\bigg)^\frac{1}{2} \\
&\simeq \pm \frac{\tilde{N} \lambda^\frac{1}{2}}{|r k_r|}\pm \mathrm{sign}(k_r)\frac{i}{2}\frac{N_\theta^2 F}{\tilde{N} \lambda^\frac{1}{2}} \, .
\end{split}
\eeq
Expanding equation \ref{eqB28} to first order in thermal diffusivity and baroclinicity,
\beq
\begin{split}
\omega^2 & \simeq \frac{\lambda}{k_r^2r^2}\bigg[N^2_T\bigg(1+\frac{i\kappa k_r^2}{\omega}\bigg)^{-1}+N_\mu^2\bigg] +\frac{i}{k_rr}N_\theta^2 F\\
&\simeq \frac{N^2\lambda}{k_r^2r^2}-i\frac{N_T^2\lambda}{k_r^2r^2}\frac{\kappa k_r^2}{\omega}+\frac{i}{k_rr}N_\theta^2 F \, .
\end{split}
\eeq
Substituting the real part of equation \ref{eqB29} for $\omega$
\beq
\begin{split}
\omega^2 &\simeq \frac{N^2\lambda}{k_r^2r^2}\mp i\frac{N_T^2\lambda}{k_r^2r^2}\frac{|k_rr|}{N \lambda^\frac{1}{2}} \kappa k_r^2+\frac{i}{k_rr}N_\theta^2 F
\end{split}
\eeq
hence
\beq
\begin{split}
\omega &\simeq \pm \frac{N\lambda^\frac{1}{2}}{r k_r}-\frac{i}{2}\frac{N_T^2}{N^2}\kappa k_r^2\pm \frac{i}{2}\mathrm{sign}(k_r)\frac{N_\theta^2 F}{N\lambda^\frac{1}{2}} \, .
\end{split}
\eeq
So the growth rate is
\beq
\label{tradgrow}
\gamma=\mathrm{Im}(\omega)=-\frac{1}{2}\frac{N_T^2}{N^2}\kappa k_r^2\pm \frac{1}{2}\mathrm{sign}(k_r)\frac{N_\theta^2 F}{N \lambda^\frac{1}{2}} \, .
\eeq
The first term is always a damping term caused by thermal diffusion, while the second term accounts for baroclinicity and can cause either growth or damping. For the typical case in stars where $q<0$ such that $N_\theta^2 < 0$, only baroclinic modes with negative $F k_r$ can grow.

In the limit of rapid thermal diffusion, i.e. $\kappa\rightarrow\infty$, we have $\tilde{N}^2\sim N_T^2\omega/i\kappa k_r^2+N_{\mu}^2$ and $\tilde{N_\theta}^2\sim N_\theta^2\omega/i\kappa k_r^2$. Then equation \ref{eqB28} reduces to
\beq
\omega^2=\frac{N_{\mu}^2\lambda}{k_r^2r^2}+\bigg(\frac{N_\theta^2 F}{k_rr}\frac{1}{\kappa k_r^2}-\frac{iN_T^2\lambda}{\kappa k_r^4 r^2}\bigg)\omega \, .
\eeq
This is a quadratic equation of the form $\omega^2-B\omega-C^2=0$, and in the limit of rapid thermal diffusion $B^2 \ll C$. Then we have
\beq
\omega \simeq \pm C+\frac{B}{2} \simeq \pm\frac{(N_\mu^2\lambda)^\frac{1}{2}}{|k_rr|} + \frac{N_\theta^2 F}{2k_rr}\frac{1}{\kappa k_r^2} - \frac{iN_T^2\lambda}{2\kappa k_r^4 r^2} \, .
\eeq
So the growth rate is
\beq
\gamma = \mathrm{Im}(\omega)= \frac{F}{2 r k_r} \frac{N_\theta^2}{\kappa k_r^2} - \frac{\lambda}{2 r^2 k_r^2} \frac{N_T^2}{\kappa k_r^2} \, .
\eeq
Baroclinic modes can still grow if $N_\theta^2 r k_r \gtrsim N_T^2$, although their growth rate is strongly suppressed by the effect of thermal diffusion.  Growth requires large wavenumbers $r k_r \gtrsim N_T^2/\Omega^2$, which is typically an enormous number in stellar interiors, so the growth rate will be very small. In fact, the required wavenumbers can be so large that molecular viscosity can become important. Hence, in many situations where thermal diffusion is important, the baroclinic instability will be quenched or grow too slowly to be relevant. 

\subsection{Geostrophic Approximation}
\label{appendixb2}

We now examine the baroclinic instability using a geostrophic approximation. For this calculation, we analyze the oscillation modes locally, in a Cartesian box at colatitude $\theta$. We adopt a coordinate system with $\hat{x}$ in the $\phi$ direction, $\hat{y}$ in the $\theta$ direction, and $\hat{z}$ in the $r$ direction. The perturbations variables are assumed to have dependence
\beq
\delta Q\propto\exp \big[ i(k_x x+k_y y+k_zz-\omega t) \big] \, .
\eeq
We have the equations of motion
\beq
\label{B40}
fv_y=\frac{1}{\rho}\frac{\partial}{\partial x}\delta P
\eeq
\beq
\label{B41}
fv_x=-\frac{1}{\rho}\frac{\partial}{\partial y}\delta P
\eeq
\beq
\label{B42}
g\delta \rho=-\frac{\partial}{\partial z}\delta P \, ,
\eeq
where $f = 2\Omega\cos \theta$. Similar to the traditional approximation, the geostrophic approximation uses $v_z \ll v_x,v_y$ and $\Omega \ll N$ to drop Coriolis terms in equations \ref{B40} and \ref{B42}. The continuity equation again uses an incompressible approximation to become
\beq
\label{B43}
\frac{\partial v_x}{\partial x}+\frac{\partial v_y}{\partial y}+\frac{\partial v_z}{\partial z}=0 \, .
\eeq
We take the time derivative of the energy equation \eqref{EE}, applying incompressible approximation, i.e. $c_\mathrm{s}\rightarrow \infty$ to find
\beq
\label{B44}
-i\omega \frac{\delta\rho}{\rho}g=\tilde{N}^2v_z+\tilde{N}_\theta^2v_y \, .
\eeq

Equations \ref{B40}-\ref{B44} can be combined to yield a geostrophic dispersion relation. Combining equations \ref{B42} and \ref{B43},
\beq
\label{B45}
i\omega\frac{1}{\rho}\frac{\partial}{\partial z}\delta P=\tilde{N}^2v_z+\tilde{N}_\theta^2v_y \, .
\eeq
Applying the WKB approximation, i.e. $k_x,k_y,k_z\gg r,H$, we take the $z$ derivative of equation \ref{B45}:
\beq
\begin{split}
&-i\omega\frac{1}{\rho}k_z^2\delta P\\
=&\tilde{N}^2\frac{\partial v_z}{\partial z}+\tilde{N}_\theta^2\frac{\partial v_y}{\partial z}\\
=&-\tilde{N}^2\bigg(\frac{\partial v_x}{\partial x}+\frac{\partial v_y}{\partial y}\bigg)+\tilde{N}_\theta^2\frac{\partial v_y}{\partial z}\\
=&-\tilde{N}^2\bigg[\frac{\partial}{\partial x}\bigg(-\frac{1}{\rho f}\frac{\partial}{\partial y}\delta P\bigg)+\frac{\partial} {\partial y}\bigg(\frac{1}{\rho f}\frac{\partial}{\partial x}\delta P\bigg) \bigg] \\
&\qquad\qquad+\tilde{N}_\theta^2\frac{\partial}{\partial z}\bigg(\frac{1}{\rho f}\frac{\partial}{\partial x}\delta P\bigg)\\
=&-i\tilde{N}^2\frac{\partial} {\partial y}\bigg(\frac{1}{\rho f}\bigg)k_x\delta P-\tilde{N}_\theta^2\frac{1}{\rho f}k_xk_z\delta P \, ,
\end{split}
\eeq
hence
\beq
\omega=\tilde{N}^2\rho\frac{\partial} {\partial y}\bigg(\frac{1}{\rho f}\bigg)\frac{k_x}{k_z^2}-i\tilde{N}_\theta^2\frac{1}{f}\frac{k_x}{k_z} \, .
\eeq
Note that in our thin box centered at $(r_0,\theta_0$), $y = r_0(\theta-\theta_0)$, so that
\beq
\frac{\partial \rho}{\partial y} =-\frac{1}{r_0}\frac{\partial \rho}{\partial \theta}=\frac{\rho}{g} N_\theta^2
\eeq
and
\beq
\frac{\partial f}{\partial y}=-\frac{1}{r_0}\frac{\partial f}{\partial \theta}=\frac{2\Omega}{r_0} \sin\theta_0 \, .
\eeq
Therefore
\beq
\omega=-\frac{k_x}{k_z^2}\tilde{N}^2\bigg(\frac{1}{f^2}\frac{2\Omega}{r_0}\sin\theta_0+\frac{1}{fg}N_\theta^2\bigg)-i\tilde{N}_\theta^2\frac{1}{f}\frac{k_x}{k_z} \, .
\eeq
Substituting the expression for $f$, and assuming $N_\theta^2 r_0\ll g$, we have
\beq
\label{B53}
\omega=-\frac{k_x}{k_z}\frac{1}{k_z r_0}\frac{\tilde{N}^2}{2\Omega}\frac{\sin\theta_0}{\cos^2\theta_0}-i\frac{\tilde{N}_\theta^2}{2\Omega\cos\theta_0}\frac{k_x}{k_z} \, .
\eeq
This is the local dispersion relation for geostrophic modes in the presence of large buoyancy and baroclinicity. Note that only retrograde Rossby modes exist, as normal gravity modes have been filtered out by the geostrophic approximation. Despite prior assertions to the contrary \citep{spruit:83,zahn:92}, the baroclinic term in equation \ref{B53} can cause local growth of baroclinic modes in the geostrophic approximation. Just as in the traditional approximation, however, only modes propagating in one direction (one sign of $k_z$) will be able to grow. Standing modes composed of waves traveling in both directions will therefore not be able to grow, as discussed in the main text.

Considering now the case with small thermal diffusion and baroclinicity, to zeroth order the wave frequency is
\beq
\label{B54}
\omega=-\frac{k_x}{k_z}\frac{1}{k_z r_0}\frac{N^2}{2\Omega}\frac{\sin\theta_0}{\cos^2\theta_0} \, .
\eeq
Expanding equation \ref{B53} in terms of thermal diffusivity, 
\beq
\begin{split}
\omega &\simeq -\frac{k_x}{k_z} \frac{1}{2\Omega k_z r_0} \frac{\sin\theta_0}{\cos^2\theta_0} \bigg[ N_T^2\bigg(1+\frac{i\kappa k^2}{\omega}\bigg)^{-1}+N_\mu^2 \bigg]\\
&\qquad\qquad-\frac{i}{2\Omega\cos\theta_0}\frac{k_x}{k_z}N_\theta \\
&\simeq -\frac{k_x}{k_z} \frac{N^2}{2\Omega k_z r_0} \frac{\sin\theta_0}{\cos^2\theta_0}+i\bigg(-\frac{N_T^2}{N^2}\kappa k^2+\frac{N_\theta^2}{2\Omega\cos\theta_0}\frac{k_x}{k_z}\bigg) \, .
\end{split}
\eeq
The growth rate is thus
\beq
\label{geogrow}
\gamma=\mathrm{Im}(\omega)=-\frac{N_T^2}{N^2}\kappa k^2+\frac{N_\theta^2}{2\Omega\cos\theta_0}\frac{k_x}{k_z}
\eeq
Just as our result for the traditional approximation, baroclinic modes can only grow for sufficiently small wavenumbers such that the baroclinic term outweighs the thermal damping term. We also note that Rossby modes in the traditional approximation have $F \sim \lambda$. For these modes, the traditional approximation growth rate equation \ref{tradgrow} can be rearranged to be identical to equation \ref{geogrow}, modulo factors of order unity.

When thermal diffusion is large and there is a finite composition gradient, we find
\beq
\label{geogrowdiff}
\omega \simeq a N_\mu^2 - i \frac{a^2 N_\mu^2 N_T^2}{k^2 \kappa} - i \frac{a^2 k_x N_\mu^4 N_\theta^2}{2 \Omega \cos \theta_0 k_z (k^2 \kappa)^2} \, ,
\eeq
where $a = -k_x \sin \theta_0/(2 r_0 k_z^2 \Omega \cos^2 \theta_0)$. Equation \ref{geogrowdiff} can be rearranged to show that growth requires $q \Omega \gtrsim k^2 \kappa (k_z/k_x) N_T^2/N_\mu^2$. In real stars, growth often requires wavenumbers small enough that the large thermal diffusion approximation is invalidated, i.e., there are often no growing modes when thermal diffusion is large. 

\section{Local Analysis of Tayler Instability}\label{appendixc}

Here we derive the dispersion relation and growth rate of unstable modes of Tayler instability.

\subsection{Dispersion relation}\label{appendixc1}
Here we give a detailed discussion about the dispersion relation of Tayler instability in spherical coordinates. This generalizes the results by \cite{tayler:73,spruit:99,zahn:07} in which only the dispersion relation near the pole is considered. Similar to preceding sections, we assume all perturbation variables $\delta Q$ have short radial wavelength due to strong buoyancy, with time/spatial dependence
\begin{align}
\delta Q \propto \exp\big[ i(k_rr&+m\phi+l\theta-\omega t) \big] \, \\
k_rr&\gg l,m \, 
\end{align}
i.e., we make a WKB approximation. The eigenvalue of Laplacian can be simplified to
\beq
k_r^2+\frac{2}{r}k_r+\frac{l^2}{r^2}-\cot\theta\frac{l}{r^2}+\frac{m^2}{r^2\sin^2{\theta}}\approx k_r^2 
\eeq
As above, we make an incompressible approximation, $c_\mathrm{s}\rightarrow \infty$. Here, however, we ignore the baroclinic term because we shall see that thermal diffusion is generally large at the short wavelengths characteristic of the Tayler instability. The energy equation \ref{EE} then reduces to
\beq
\label{energy}
-g\frac{\delta\rho}{\rho}=-\bigg(N_t^2\bigg(1+i\frac{\kappa k_r^2}{\omega}\bigg)^{-1}+N_\mu^2\bigg)\xi_r \, .
\eeq

Including magnetic forces in the MHD limit, the perturbed induction equation is
\beq
-i\omega \vec{b}=\nabla\times(-i\omega \vec{\xi}\times\vec{B})-\eta k_r^2 \vec{b} \, .
\eeq
In differentially rotating stars, the azimuthal magnetic field is generated by winding up a weak poloidal magnetic field. The instability requires the azimuthal field to become much larger than the poloidal seed field, so we assume a background (unperturbed) field of 
\beq
\label{B}
\vec{B}\approx B_\phi \hat{\phi} \approx B_0\sin\theta \hat{\phi} \, .
\eeq
Here, we have assumed latitudinal dependence proportional to $\sin \theta$ for the background field, which arises from our assumption of rotation constant on spherical shells, i.e., $\partial v_{\rm rot} \partial r = q \Omega \sin \theta$. Substituting equation \ref{B} into the induction equation, we get
\beq
\begin{split}
&\bigg(1+\frac{i\eta k_r^2}{\omega}\bigg)\vec{b}=\nabla\times(\vec{\xi}\times B_0\sin\theta \, \hat{\phi})\\
&=\vec{\xi}(\nabla\cdot B_0\sin\theta \hat{\phi})-B_0\sin\theta\hat{\phi}(\nabla\cdot\vec{\xi})\\
&\quad\quad+(\vec{\xi}\cdot\nabla)B_0\sin\theta\hat{\phi}-(B_0\sin\theta\hat{\phi}\cdot\nabla)\vec{\xi}\\
&=im\frac{B_0}{r}\vec{\xi}-\frac{B_0\sin\theta \xi_r}{r}\hat{\phi}
\end{split}
\eeq
where we applied incompressible approximation, i.e. $\nabla\cdot\vec{\xi}=0$. Due to the strong buoyancy force in radial direction, $\xi_r$ is small thus the second term can be neglected, then
\beq
\label{b}
\vec{b}=(1+i\eta k_r^2/\omega)^{-1}im\frac{B_0}{r}\vec{\xi}
\eeq

Now we calculate the perturbed Lorentz force, to first order
\beq
\vec{L}=\frac{1}{4\pi\rho} \big[ (\nabla\times\vec{b})\times\vec{B}+(\nabla\times\vec{B})\times\vec{b} \big]
\eeq
By \eqref{B} and \eqref{b}, we have
\beq
\begin{split}
\nabla\times\vec{b}&\approx(1+i\eta k_r^2/\omega)^{-1}im\frac{B_0}{r}\nabla\times\vec{\xi}\\
&=(1+i\eta k_r^2/\omega)^{-1}im\frac{B_0}{r}\\
\times \bigg[ \bigg(i\frac{l}{r}\xi_\phi&-i\frac{m}{r\sin\theta}\xi_\theta\bigg)\hat{r}+\bigg(i\frac{m}{r\sin\theta}\xi_r-ik_r\xi_\phi\bigg)\hat{\theta} \\
&+\bigg(ik_r\xi_\theta - i \frac{l}{r} \xi_r \bigg)\hat{\phi} \bigg] \, \\
\end{split}
\eeq
and so
\beq
\begin{split}
(\nabla\times&\vec{b})\times\vec{B}=(1+i\eta k_r^2/\omega)^{-1}\frac{mB_0^2}{r^2}\sin\theta \\
&\times \bigg[ \bigg(l\xi_\phi-\frac{m}{\sin\theta}\xi_\theta\bigg)\hat{\theta}-\bigg(\frac{m}{\sin\theta}\xi_r-k_rr\xi_\phi\bigg)\hat{r} \bigg]
\end{split}
\eeq
and
\beq
\begin{split}
 (\nabla\times&\vec{B})\times\vec{b}=(1+i\eta k_r^2/\omega)^{-1}\frac{mB_0^2}{r^2}\sin\theta \\
& \times \big[ -i\xi_\phi\hat{r} - 2i\xi_\phi\cot\theta\hat{\theta} + (2i\cot\theta\xi_\theta+i\xi_r)\hat{\phi} \big]
\end{split}
\eeq
where we applied the WKB approximation, hence the Lorentz force is
\begin{align}
L_r&=\frac{m\omega_\mathrm{A}^2}{1+i\eta k_r^2/\omega}(k_rr\sin\theta\xi_\phi-i\sin\theta\xi_\phi-m\xi_r)\\
L_\theta&=\frac{m\omega_\mathrm{A}^2}{1+i\eta k_r^2/\omega}(l\sin\theta\xi_\phi-m\xi_\theta-2i\cos\theta\xi_\phi)\\
L_\phi&=\frac{m\omega_\mathrm{A}^2}{1+i\eta k_r^2/\omega}(2i\cos\theta\xi_\theta+i\sin\theta\xi_r)
\end{align}
where $\omega_\mathrm{A}=B_0/\sqrt{4\pi\rho}r$ is the Alfv\'{e}n frequency.

We also have the perturbed equation of motion
\beq
\frac{1}{\rho}\nabla\delta P-\omega^2\vec{\xi}+g\frac{\delta \rho}{\rho}-2i\omega \Omega\hat{z}\times\vec{\xi}-\vec{L}=0
\eeq
Substituting the expressions for Lorentz force and \eqref{energy} into the equation of motion gives
\beq
\label{motion_r}
\begin{split}
ik\frac{\delta P}{\rho}-&\frac{(k_rr\sin\theta)m\omega_\mathrm{A}^2}{1+i\eta k_r^2/\omega}\xi_\phi\\
+\bigg(-\omega^2&+\frac{m^2\omega_\mathrm{A}^2}{1+i\eta k_r^2/\omega}+\frac{N_t^2}{1+i\kappa k_r^2/\omega}+N_\mu^2\bigg)\xi_r\\
+&\bigg(i\frac{m\omega_\mathrm{A}^2\sin\theta}{1+i\eta k_r^2/\omega}+2i\Omega\omega\sin\theta\bigg)\xi_\phi=0
\end{split}
\eeq
\beq
\label{motion_theta}
\begin{split}
i\frac{l}{r}\frac{\delta P}{\rho}+\bigg(-\omega^2+\frac{m^2\omega_\mathrm{A}^2}{1+i\eta k_r^2/\omega}\bigg)\xi_\theta&+\\
\bigg(-\frac{(l \sin\theta)m\omega_\mathrm{A}^2}{1+i\eta k_r^2/\omega}+\frac{2 i m\omega_\mathrm{A}^2\cos\theta}{1+i\eta k_r^2/\omega}&+2i\Omega\omega\cos\theta\bigg)\xi_\phi=0
\end{split}
\eeq
\beq
\label{motion_phi}
\begin{split}
i\frac{m}{r\sin\theta}\frac{\delta P}{\rho}&-\bigg(i\frac{2m\omega_\mathrm{A}^2\cos\theta}{1+i\eta k_r^2/\omega}+2i\Omega\omega\cos\theta\bigg)\xi_\theta\\
-\omega^2\xi_\phi&-\bigg(i\frac{m\omega_\mathrm{A}^2\sin\theta}{1+i\eta k_r^2/\omega}+2i\Omega\omega\sin\theta\bigg)\xi_r=0 \, .
\end{split}
\eeq

We now define
\begin{align}
\xi_0&\equiv\frac{\delta P}{\rho}+i\frac{r\sin\theta m\omega_\mathrm{A}^2}{1+i\eta k_r^2/\omega}\xi_\phi\\
A&\equiv-\omega^2+\frac{m^2\omega_\mathrm{A}^2}{1+i\eta k_r^2/\omega}\\
B&\equiv i\frac{2m\omega_\mathrm{A}^2\cos\theta}{1+i\eta k_r^2/\omega}+2i\Omega\omega\cos\theta\\
C&\equiv i\frac{m\omega_\mathrm{A}^2\sin\theta}{1+i\eta k_r^2/\omega}+2i\Omega\omega\sin\theta\\
D&\equiv\frac{N_t^2}{1+i\kappa k_r^2/\omega}+N_\mu^2 \, .
\end{align}
The continuity equation and equations \eqref{motion_r}-\eqref{motion_phi} reduce to
\begin{align}
k_r\xi_r+\frac{l}{r}\xi_\theta+\frac{m}{r\sin\theta}\xi_\phi&=0\\
i\frac{l}{r}\xi_0+A\xi_\theta+B\xi_\phi&=0\\
i\frac{m}{r\sin\theta}\xi_0-C\xi_r-B\xi_\theta+A\xi_\phi&=0\\
ik_r\xi_0+(A+D)\xi_r+C\xi_\phi&=0 \, .
\end{align}

The dispersion relation is given by the determinant of the linear equations, i.e.,
\beq
\begin{split}
&\left| \begin{array}
{cccc} 0 & k_r & \frac{l}{r} & \frac{m}{r\sin\theta} \\ 
i\frac{l}{r} & 0 & A & B \\
i\frac{m}{r\sin\theta} & -C & -B & A \\
ik_r & A+D & 0 & C 
\end{array} \right|=\\
\bigg[ \bigg(\frac{l}{r}\bigg)^2&+\bigg(\frac{m}{r\sin\theta}\bigg)^2 \bigg] A(A+D)+k_r^2(A^2+B^2)\\
&-k_r\frac{l}{r}BC=0
\end{split}
\eeq
For Tayler modes, we shall see that $A\sim B \sim C \ll D$, so we further define
\beq
\tilde{D}\equiv\bigg(\frac{l}{k_rr}\bigg)^2D \, .
\eeq
Considering that $D\gg A$
\beq
k_r^2(A^2+B^2+A\tilde{D})-k_r\frac{l}{r}BC=0
\eeq
From our WKB approximation in the radial direction, $k_rr\gg l$, thus we can neglect the last term, and the dispersion relation reduces to
\beq
\label{dispersion}
A^2+B^2+A\tilde{D}=0 \, .
\eeq
Next, we define the following dimensionless variables (the same definitions used in \citealt{zahn:07}) as
\beq
\begin{split}
\tilde{\omega}=\frac{\omega}{\omega_\mathrm{A}},\;\tilde{\Omega}=\frac{\Omega}{\omega_\mathrm{A}},\;k=\frac{\kappa k_r^2}{\omega_\mathrm{A}},\;h=\frac{\eta k_r^2}{\omega_\mathrm{A}}\\
A_t=\bigg(\frac{l}{k_rr}\bigg)^2\frac{N_t^2}{\omega_\mathrm{A}^2},\;A_\mu=\bigg(\frac{l}{k_rr}\bigg)^2\frac{N_\mu^2}{\omega_\mathrm{A}^2},\;
\end{split}
\eeq
Substituting the expressions for $A,B$ and $\tilde{N}$, and after some algebra, we arrive at the dispersion relation
\beq
\label{Zahn}
\begin{split}
&\tilde{\omega}^6 -\tilde{\omega}^4(4\tilde{\Omega}^2\cos^2\theta+A_t+A_\mu+2m^2+2hk+h^2)\\
-&\tilde{\omega}^3\cdot 8 m \tilde{\Omega} + \tilde{\omega}^2 \big[ m^2A_t+m^2A_\mu+m^2-4m^2\cos^2\theta\\
+&2hk(4\tilde{\Omega}^2\cos^2\theta+A_\mu+m^2)+h^2 (4\tilde{\Omega}^2\cos^2\theta+A_t+A_\mu) \big]\\
+&\tilde{\omega}\cdot8mkh\tilde{\Omega}\cos^2\theta-hkm^2A_\mu\\
+&i\tilde{\omega}^5(2h+k) -i\tilde{\omega}^3 \big[ k(4\tilde{\Omega}^2\cos^2\theta+A_\mu+2m^2)\\
+&2h(4\tilde{\Omega}^2\cos^2\theta+A_t+A_\mu+m^2)+h^2k \big]\\
-&i\tilde{\omega}^2\cdot 8 m (k+h)\tilde{\Omega}\cos^2\theta \\
+&i\tilde{\omega} \big[ k(m^2A_\mu+m^2-4m^2\cos^2\theta)\\
+&hm^2(A_t+A_\mu)+h^2k(4\tilde{\Omega}^2\cos^2\theta+A_\mu) \big]\\
=&0
\end{split}
\eeq
which is exactly a generation of the dispersion relation near the pole given by \cite{zahn:07}, whose result is only valid near the pole while our result works at arbitary latitude.

\subsection{Unstable modes}\label{appendixc2}

Equation \ref{Zahn} is a little lengthy thus difficult to analyze, so we start with the more compact equation \ref{dispersion}. Using the same notation as \cite{spruit:99}, after a little algebra, equation \ref{dispersion} reduces to
\beq
\begin{split}
&\bigg[ \omega-\frac{m^2\omega^2_A}{\omega+i\eta k_r^2}-\bigg(\frac{l}{k_rr}\bigg)^2\frac{N^2_t}{\omega+i\kappa k_r^2}-\bigg(\frac{l}{k_rr}\bigg)^2\frac{N_\mu^2}{\omega} \bigg]\\
\times& \big[ \omega(\omega+i\eta k_r^2)-m^2\omega_A^2 \big] - \bigg(2\Omega\cos\theta+\frac{2m\omega^2_A\cos\theta}{\omega+i\eta k_r^2}\bigg)\\
&\times \big[ 2\Omega\cos\theta(\omega+i\eta k_r^2) + 2m\omega_A^2\cos\theta \big] = 0
\end{split}
\eeq
This is our version of equation (A8) from \cite{spruit:99}. Note that we have set $p=1$ here, which is appropriate for a background magnetic field with $B \propto \sin \theta$ as expected for a field generated by winding a radial seed field. 

For $\Omega\gg\omega_A$, we expect $\omega\sim\omega^2_A/\Omega$ for both real and imaginary parts, so we define
\beq
\omega\equiv\alpha\frac{\omega_A^2}{\Omega},\quad H\equiv\frac{\eta \Omega}{\omega^2_A},\quad K\equiv\frac{\kappa\Omega}{\omega_A^2}
\eeq
and
\beq
n^2=\bigg(\frac{l}{k_rr}\bigg)^2\frac{N_\mu^2}{\omega_A^2}
\eeq
When the thermal diffusion is large, which is often the case in stars, we can take the limit $K \rightarrow \infty$. Note that the limit $K \rightarrow \infty$ is the same as the limit $k \rightarrow 0$, $N = N_\mu$ considered by \cite{spruit:99}. We further limit our analysis to slow modes, i.e., we take the limit $\omega \ll \Omega$, which eliminates $r$ modes and gravity modes. The dispersion relation then reduces to
\beq
\label{A18}
m^2 \bigg[ m^2\alpha+n^2(\alpha+iHk_r^2) \bigg]-4\big[(\alpha+iHk_r^2)+m\big]^2\alpha\cos^2\theta=0
\eeq
This is our version of equation (A18) from \cite{spruit:99} for the case $p=1$. The latitudinal dependence arises only through the factor of $\cos^2 \theta$ in the last term.

A growing mode requires equation \eqref{A18} to have a solution with a positive imaginary part, which straightforwardly excludes $m=0$ modes from our interest. Though the cubic complex equation \eqref{A18} is in general difficult to solve, it is easy to see that in the case with large magnetic diffusivity, i.e. $Hk_r^2\gg1$, the solutions of $\alpha$ must have a large negative imaginary part of $\sim -Hk_r^2$ to make the LHS vanish. Hence, large magnetic diffusion or very large wavenumber $k_r^2$ will inevitably lead to damped modes.

Let us fix the azimuthal wave number $m$ and the polar angle $\theta$ for now. We start with some $n^2, Hk_r^2$ so that the solution of \eqref{A18} is a damped mode, which is always possible by setting $Hk_r^2$ large enough. By continuously changing the parameters $n^2, Hk_r^2$, the coefficients and solutions of the cubic equation vary continuously. The boundary between damped and growing modes is defined by some set of $n^2, Hk_r^2$ that makes $\alpha$ purely real. Hence, requiring a real solution of \eqref{A18} defines an instability criterion, with mode growth only possible on one side of this boundary, and no growing modes if there is no solution for a purely real $\alpha$. By setting $\alpha$ real, we can separate \eqref{A18} into its real and imaginary parts:
\begin{align}
\begin{split}
\text{Re:}\qquad& m^2(m^2+n^2)-4\cos^2\theta(\alpha+m)^2\\
+4&\cos^2\theta H^2k_r^4=0
\end{split}\\
\text{Im:}\qquad&m^2 n^2-8\cos^2\theta\alpha(\alpha+m)=0
\end{align}
Eliminating $\alpha$ from the equations above, we have
\begin{align}
\label{C41}
&2 m^4 + m^2 n^2 - 4 m^2\cos^2\theta\bigg(1\pm\sqrt{1+\frac{1}{2}\frac{n^2}{\cos^2\theta}}\bigg) \nonumber \\ & + 8\cos^2\theta H^2k_r^4=0
\end{align}

Equation \eqref{C41} must be satisfied for \eqref{A18} to have a real solution such that growing modes exist. Since $\cos^2\theta H^2k_r^4>0$, we must have
\beq
2m^4+ m^2 n^2 < 4 m^2\cos^2\theta\bigg(1\pm\sqrt{1+\frac{1}{2}\frac{n^2}{\cos^2\theta}}\bigg)
\eeq
Defining $t^2 = n^2/\cos^2\theta$, we have
\beq
\frac{m^2}{2\cos^2\theta}<-\frac{t^2}{4}+1\pm\sqrt{1+\frac{1}{2}t^2}
\eeq
Since $t^2>0$, the maximum of the right-hand side occurs when the positive sign is chosen with $t\rightarrow0$, where the maximum value $\rightarrow2$. Thus, the instability can only grow if
\beq
m^2<4\cos^2\theta
\eeq
This is only possible for $m=1$ modes, with
\beq
\cos^2\theta>\frac{1}{4}
\eeq
Hence the instability can only occur at high latitudes where $\theta<\pi/3$ or $\theta>2\pi/3$.

For $m=1$ and $\cos \theta =1$, instability requires 
\beq
\frac{1}{2} - \frac{t^2}{4} + \sqrt{1 + t^2/2} > 0 \, ,
\eeq
which requires
\beq
\frac{1}{2} - \frac{t^2}{4} + \sqrt{1 + t^2/2} > 0 \, ,
\eeq
and it is easy to show this requires 
\beq 
\bigg( \frac{l N_\mu}{r k_r \omega_{\rm A}} \bigg)^2 < 6 + 4 \sqrt{3} \, .
\eeq
So we see that growth requires wavelengths not much larger than the Tayler instability length scale $\sim (\omega_{\rm A}/N_\mu) r$. Looking back at equation \ref{C41}, we see that growing modes also require
\beq
m^2 \big(1 + \sqrt{1+n^2/(2 \cos^2 \theta)} \bigg) > 2 (H k_r^2)^2 \, .
\eeq
From above, we found that only modes $m=1$, with $n$ less than a few, and with $\cos \theta > 1/2$ can grow. Then the left-hand side of the above equation is of order unity for growing modes, and growth additionally requires $Hs^2 \lesssim 1$. Growing modes must therefore have wavelengths $\gtrsim \sqrt{\eta \Omega}/\omega_{\rm A}$. Combining this criterion with the maximum wavelength criterion from above yields the criterion of \cite{spruit:02} for Tayler instability that
\beq
\omega_{\rm A}^4 \gtrsim \frac{N_\mu^2 \Omega \eta}{r^2} \, .
\eeq

We do not attempt to solve the cubic equation for mode growth rates as was done by \cite{zahn:07}. However, we note that when growth occurs, the fastest growing modes are those with $Hs^2$ of order unity. The wavelength of the fastest growing modes is similar to (but slightly larger than) those on the verge of stability, with scales of $\sim \sqrt{\eta \Omega}/\omega_{\rm A}$. When $\omega_{\rm A} \sim \omega_c = (N_{\rm eff}^2 \Omega \eta/r^2)^{1/4}$, as often happens in our models, the fastest growing modes have $r k_r \sim N_{\rm eff}/\omega_{\rm A}$.

\end{document}